\documentclass[12pt]{article}
\usepackage{graphicx}
\usepackage{amsmath, amssymb}

\newcommand{\half}{{\frac{1}{2}}}
\newcommand{\ket}[1]{\,\left|\,{#1}\right\rangle}
\newcommand{\mbf}[1]{\mathbf{#1}}

\renewcommand{\bar}[1]{\overline{#1}}

\def\Dslash{\raise.15ex\hbox{/}\kern-.7em D}
\def\Pslash{\raise.15ex\hbox{/}\kern-.7em P}

\begin {document}
\begin{flushright}
{\small
SLAC--PUB--13107\\
January 2008}
\end{flushright}

\vspace{20pt}

\centerline{\LARGE \bf AdS/CFT  and Light-Front QCD}

\vspace{15pt}

\centerline{\bf {
Stanley J. Brodsky$^{a}$
%\footnote{Electronic address:sjbth@slac.stanford.edu}
 and
Guy F. de T\'eramond$^{b}$}}
%\footnote{Electronic address:gdt@asterix.crnet.cr}

\vspace{10pt}

{\centerline {$^{a}${Stanford Linear Accelerator Center, 
Stanford University, Stanford, CA 94309, USA}}

\vspace{2pt}

{\centerline {$^{b}${Centre de Physique Th\'eorique,  Ecole Polytechnique, 91128 Palaiseau, France}}

 \vspace{15pt}

\begin{abstract}

The AdS/CFT correspondence between string theory in AdS space and
conformal field theories in physical space-time leads  to an
analytic, semi-classical model for strongly-coupled QCD which has
scale invariance and dimensional counting at short distances and
color confinement at large distances.  The
AdS/CFT correspondence also provides insights into the inherently
nonperturbative aspects of QCD such as the orbital and radial
spectra of hadrons and the form of hadronic wavefunctions.  In
particular, we show that there is an exact correspondence between
the fifth-dimensional coordinate of AdS space $z$ and a specific
impact variable $\zeta$ which measures the separation of the quark
and gluonic constituents within the hadron in ordinary space-time.
This connection leads to AdS/CFT predictions for the analytic form of the
frame-independent light-front wavefunctions (LFWFs) of mesons and baryons,
the fundamental entities which encode hadron properties. The LFWFs in turn predict
decay constants and spin correlations,  as well as dynamical quantities such as form factors, structure functions, generalized parton distributions, and
exclusive scattering amplitudes.
Relativistic light-front equations in ordinary space-time are found
which reproduce the results obtained using the fifth-dimensional
theory and have remarkable algebraic structures and
integrability properties. As specific
examples we describe the
behavior of the pion form factor  in the space and
time-like regions and determine the Dirac nucleon form factors in the space-like region.
An extension to nonzero quark mass is used to determine hadronic
distribution amplitudes of all mesons, heavy and light. We compare our results with the moments of the distribution amplitudes  which have recently been computed from lattice gauge theory.

\end{abstract}

\vspace{5pt}

\begin{center}
{\it Two lectures presented at  the International School of Subnuclear Physics\\
Searching for the `Totally Unexpected' in the LHC Era\\
Erice, Sicily, August 29 -- September 7, 2007
 }
\end{center}

\vfill

%\tableofcontents
\newpage

\section{Introduction}

Quantum Chromodynamics, the Yang-Mills local gauge field theory of
$SU(3)_C$ color symmetry provides a fundamental description of
hadron and nuclear physics in terms of quark and gluon degrees of
freedom. Yet, because of its strong coupling nature, it has been 
difficult to find analytic solutions to QCD or to make precise
predictions outside of its perturbative domain.  An important theoretical goal
is thus to find an initial approximation to QCD which is both
analytically tractable and which can be systematically improved. For
example, in quantum electrodynamics, the Coulombic Schr\"odinger and
Dirac equations provide quite accurate first approximations to
atomic bound state problems, which can then be systematically
improved using the Bethe-Salpeter formalism and correcting for
quantum fluctuations, such as the Lamb Shift and vacuum
polarization.

It was originally believed that the AdS/CFT mathematical correspondence could
only be applied to strictly conformal theories, such as
$\mathcal{N}=4$ supersymmetric Yang-Mills gauge theory.   
Conformal symmetry is broken in physical QCD by quantum effects and quark masses.
There are indications, however  both from
theory and phenomenology, that the QCD coupling is slowly varying at small momentum transfer.
In these lectures we shall discuss how conformal symmetry, plus a simple ansatz for color confinement, provides  a remarkably accurate first approximation for QCD. 

The essential element for the application of AdS/CFT to hadron physics is the indication that the QCD coupling $\alpha_s(Q^2)$ becomes large and
constant in the low momentum domain $Q \le 1$ GeV/c, thus providing a window where conformal symmetry can be applied. Solutions of the Dyson-Schwinger equations for 
the three-gluon and four-gluon couplings~\cite{vonSmekal:1997is, Zwanziger:2003cf, Alkofer:2004it, Fischer:2006vf, Epple:2006hv, Kellermann:2008iw, Alkofer:2008jy} and phenomenological
studies~\cite{Mattingly:1993ej, Brodsky:2002nb, Baldicchi:2002qm} of QCD couplings based on physical observables such as $\tau$
decay~\cite{Brodsky:1998ua}  and the Bjorken sum rule~\cite{Deur:2005cf}, show that the QCD $\beta$ function vanishes and  $\alpha_s(Q^2)$ become constant at small
virtuality; {\em i.e.}, effective charges develop an ``infrared fixed point."  Recent lattice simulations~\cite{Furui:2006py, Appelquist:2007hu} and
nonperturbative analyses~\cite{Antonov:2007mx} have also indicated an infrared fixed point for QCD.  One can understand this
physically~\cite{Brodsky:2007pt}: in a confining theory where gluons have an effective 
mass~\cite{Cornwall:1981zr} or maximal wavelength, all vacuum polarization
corrections to the gluon self-energy decouple at long wavelength; thus an infrared fixed point appears to be a natural consequence of confinement. 
Furthermore, if one considers a
semi-classical approximation to QCD with massless quarks and without
particle creation or absorption, then the resulting $\beta$ function
is zero, the coupling is constant, and the approximate theory is
scale and conformal invariant~\cite{Parisi:1972zy,Brodsky:1985ve}, allowing the mathematical tools of conformal symmetry to be applied.  One can use conformal symmetry as 
a {\it template}, systematically correcting for its nonzero $\beta$ function as well
as higher-twist effects.

One of the key consequences of conformal invariance are the dimensional counting rules~\cite{Brodsky:1974vy,Matveev:1973ra}.
The leading power fall-off of a hard exclusive process follows from the conformal scaling of the
underlying hard-scattering amplitude: $T_H \sim 1/Q^{n-4}$, where $n$ is the total number of fields (quarks, leptons, or gauge fields)
participating in the hard scattering. Thus the reaction is dominated by subprocesses and Fock states
involving the minimum number of interacting fields.
In the case of $2 \to 2$ scattering processes, this implies $ {d\sigma/ dt}(A B \to C D)
={F_{A B \to C D}(t/s)/ s^{n-2}},$ where $n = N_A + N_B + N_C +N_D$ and $n_H$ is the minimum number of constituents of $H$. The near-constancy
of the effective QCD coupling helps explain the empirical success of dimensional counting rules for the near-conformal power law fall-off of
form factors and fixed angle scaling~\cite{Brodsky:1989pv}.  For example, one sees the onset of perturbative QCD scaling behavior even for
exclusive nuclear amplitudes such as deuteron photodisintegration, here $n = 1+ 6 + 3 + 3 = 13$, $s^{11}{ d\sigma/dt}(\gamma d \to p n) \sim $
constant at fixed CM angle.

In the case of hard exclusive reactions~\cite{Lepage:1980fj}, the virtuality of the gluons exchanged in the underlying QCD process is typically much less than the momentum transfer scale $Q$, 
as  several gluons share the total momentum transfer.  Since the coupling is probed in the conformal window, this kinematic feature can explain why the measured  proton Dirac form factor scales as $Q^4 F_1(Q^2) \simeq {\rm const}$ up to $Q^2 < 35$ GeV$^2$~\cite{Diehl:2004cx} with little sign of the logarithmic running of the QCD coupling.
Thus conformal symmetry can be a useful
first approximant even for physical QCD.
The measured deuteron form factor also appears to follow the leading-twist QCD predictions at large momentum transfers in the few GeV
region~\cite{Holt:1990ze,Bochna:1998ca,Rossi:2004qm}.

Recently the Hall A collaboration at Jefferson Laboratory~\cite{Danagoulian:2007gs} has reported a significant exception to the general
empirical success of dimensional counting in fixed CM angle Compton scattering ${d\sigma / dt}(\gamma p \to \gamma p) \sim
F(\theta_{CM})/ s^8$, instead of the predicted $1/s^6$ scaling.  However, the hadron form factor $R_V(T)$, which multiplies the
$\gamma q \to \gamma q$ amplitude is found by Hall-A to scale as $1/ t^2$, in agreement with the  PQCD and AdS/CFT prediction. In addition
the timelike two-photon process $\gamma \gamma \to p \bar p$ appears to satisfy dimensional counting~\cite{Chen:2001sm,Chen:2006ug}.

Our main tool for implementing conformal symmetry will be the use of Anti-de-Sitter (AdS$_5$) space in five dimensions which provides a mathematical realization of the group $SO(4,2)$, the group of Poincare' plus conformal transformations.  
The AdS metric is
\begin{equation} \label{eq:AdSz}
ds^2 = \frac{R^2}{z^2}(\eta_{\mu \nu} dx^\mu
 dx^\nu - dz^2),
 \end{equation}
which is invariant under scale changes of the
coordinate in the fifth dimension $z \to \lambda z$ and $ x_\mu \to
\lambda x_\mu$.  Thus one can match scale transformations of the
theory in $3+1$ physical space-time to scale transformations in the
fifth dimension $z$.
The isomorphism of the group of Poincare' and conformal transformations
$SO(4,2)$ to the group of isometries of Anti-de Sitter space underlies 
the AdS/CFT
correspondence~\cite{Maldacena:1997re} between string states defined
on the 5-dimensional Anti--de Sitter (AdS) space-time and conformal
field theories in physical space-time~\cite{Gubser:1998bc, Witten:1998qj} . 
In particular, we  shall show that there is an exact correspondence between
the fifth-dimensional coordinate of AdS space $z$ and a specific
impact variable $\zeta$ which measures the separation of the quark
and gluonic constituents within the hadron in ordinary space-time.
This connection leads to AdS/CFT predictions for the analytic form of the
frame-independent light-front wavefunctions (LFWFs) of mesons and baryons,
the fundamental entities which encode hadron properties. The LFWFs in turn predict
decay constants and spin correlations,  as well as dynamical quantities such as form factors, structure functions, generalized parton distributions, and
exclusive scattering amplitudes.

Scale-changes in the physical
$3+1$ world can thus be represented by studying dynamics in a mathematical fifth dimension with the ${\rm AdS}_5$ metric. Different values of the holographic variable $z$ determine the scale of the invariant
separation between the partonic constituents. This  is illustrated in Fig. \ref{fig1}.
Hard scattering processes occur in the small-$z$ ultraviolet (UV)
region of AdS space. In particular,
the $Q \to \infty$ zero separation limit corresponds to the $z \to 0$ asymptotic boundary, where the QCD
Lagrangian is defined. 
\begin{figure}[htb]
\centering
\includegraphics[angle=0,width=10cm]{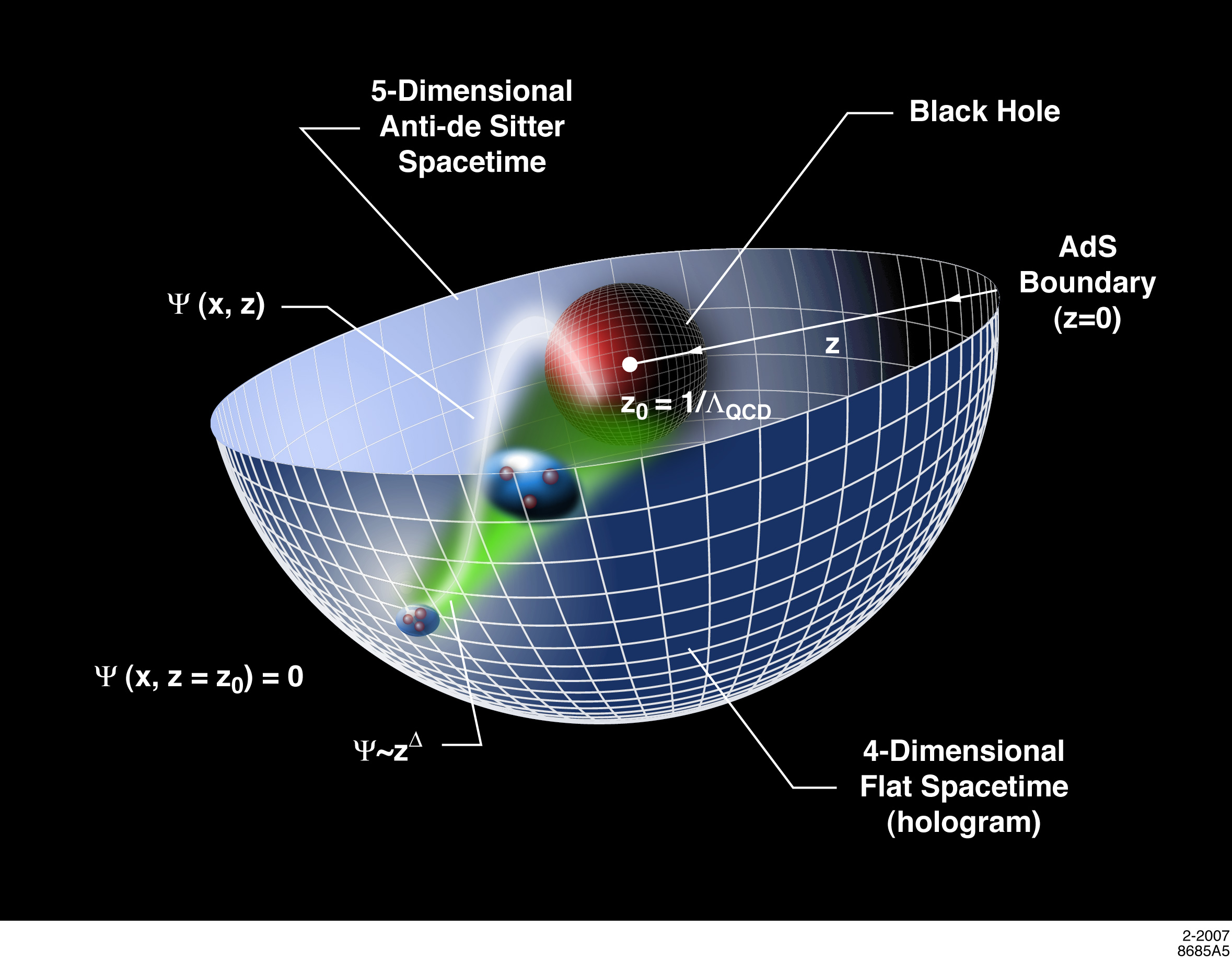}
\caption{Artist's conception of AdS/CFT.  The evolution of the
proton at different length scales is mapped into the compact AdS$_5$
dimension $z$. Dirichlet bag-like boundary condition,
 $\Psi(z)\vert_{z=z_0}=0$,  is imposed at the
 confinement radius $z = z_0 ={1/\Lambda_{\rm QCD}}$,
thus limiting interquark separations.}
\label{fig1}
\end{figure}

As shown by Polchinski
and Strassler~\cite{Polchinski:2001tt}, one can simulate confinement
by imposing boundary conditions in the holographic variable $z$. 
The infrared (IR) cut-off at $z_0 ={1/\Lambda_{\rm QCD}}$ breaks conformal
invariance,  allowing the introduction of the QCD mass scale and a spectrum of particle states.
In the hard wall model~\cite{Polchinski:2001tt} a cut-off is placed at a
finite value $z_0 = 1/\Lambda_{\rm QCD}$ and the spectrum of states is linear in the radial and angular momentum quantum numbers:
$\mathcal{M} \sim 2 n + L$. In the soft wall model a smooth infrared cutoff is chosen to model
confinement and reproduce the usual Regge behavior 
$\mathcal{M}^2 \sim n + L$~\cite{Karch:2006pv}.
The resulting models, although {\it ad hoc}, provide a
simple semi-classical approximation to QCD which has both constituent counting
rule behavior at short distances and confinement at large distances.

It is thus natural, as a useful first approximation, to use the isometries of AdS to map the local interpolating operators at the UV boundary of AdS space, into the modes propagating inside AdS. Under conformal transformations the interpolating operators transform according to their twist, and consequently the AdS isometries map the twist scaling dimensions into the AdS 
modes~\cite{Brodsky:2003px}.  
A physical hadron in four-dimensional Minkowski space has
four-momentum $P_\mu$ and 
invariant mass  given by $P_\mu P^\mu = \mathcal{M}^2$. 
The physical states in  AdS$_5$ space are represented by normalizable ``string" modes
$\Phi_P(x,z) \sim  e^{- i P\cdot x}~ \Phi(z)$,
with plane waves along the Poincar\'e coordinates and a profile
function $\Phi(z)$ along the holographic coordinate $z$, as illustrated in Fig. \ref{fig1}.
For small-$z$, $\Phi$ scales as $\Phi \sim z^\Delta$, 
where $\Delta$ is the conformal dimension of the string state, the
same dimension of the interpolating operator $\mathcal{O}$ which creates a hadron
out of the vacuum~\cite{Polchinski:2001tt}, 
$\langle P \vert \mathcal{O} \vert 0 \rangle \ne 0$.
The scale dependence of each 
string mode $\Phi(x,z)$ is thus determined by matching its behavior at $z \to 0$
with the scaling dimension of 
the corresponding hadronic state at short distances $x^2 \to 0$.
Changes in length scale are mapped to evolution in the holographic
variable $z$. The string mode  
in $z$ thus represents the extension
of the hadron wave function into the fifth dimension. The eigenvalues of normalizable modes
in AdS give the hadronic spectrum. AdS modes represent also the probability
amplitude for the distribution of quarks and gluons at a given scale.
There are also non-normalizable modes which are related to 
external currents: they propagate into the AdS  interior  and couple to
boundary QCD interpolating operators~\cite{Gubser:1998bc, Witten:1998qj}.

Following the approach described above,  a limited set of operators is  introduced to construct 
phenomenological viable five-dimensional dual holographic 
models. This simple prescription, which has been described as a ``bottom-up''
approach, has been successful in obtaining
general properties of scattering amplitudes of hadronic bound
states at strong coupling
~\cite{Polchinski:2001tt, Brodsky:2003px, Janik:1999zk, Lin:2006rf, Brower:2007qh, Penedones:2007ns}, the low-lying hadron
spectra~\cite
{Karch:2006pv, Erlich:2005qh, Boschi-Filho:2002vd,  deTeramond:2004qd, deTeramond:2005su, 
Brodsky:2006uqa, Evans:2006ea, Hong:2006ta, Colangelo:2007pt, Forkel:2007ru}, 
hadron couplings and chiral symmetry breaking~\cite{Erlich:2005qh, DaRold:2005zs, 
Hirn:2005nr, Ghoroku:2005vt, Krikun:2008tf},
quark potentials in confining
backgrounds~\cite{BoschiFilho:2005mw, Andreev:2006ct},
a description of weak hadron decays~\cite{Hambye:2005up}
and euclidean correlation functions~\cite{Schafer:2007qy}.
Geometry back-reaction in AdS may also be relevant to the 
infrared physics~\cite{Csaki:2006ji} and wall dynamics~\cite{Batell:2008zm}.
The gauge theory/gravity duality also provides a convenient framework
for the description of deep inelastic scattering structure functions at small
$x$~\cite{Polchinski:2002jw, Hatta:2007he, BallonBayona:2007rs},
a unified description of hard and soft pomeron physics~\cite{Brower:2006ea} and
gluon scattering amplitudes at strong coupling~\cite{Alday:2007hr}.

In the top-down approach, one introduces higher
dimensional branes to the ${\rm AdS}_5 \times {\rm S}^5$
background~\cite{Karch:2002sh} in order to have a theory 
of flavor. One can obtain models with massive quarks in the fundamental representation,  compute 
the hadronic spectrum, and describe chiral 
symmetry breaking in the context of higher dimensional brane constructs
~\cite{Karch:2002sh,Kruczenski:2003be, Babington:2003vm, 
Sakai:2004cn, Gursoy:2007er}.  However, a theory dual to QCD is unknown, and this ``top-down" approach 
is difficult to extend beyond theories exceedingly constrained by their
symmetries~\cite{Erdmenger:2007cm}.

As we shall discuss, there is a remarkable mapping between the AdS description of hadrons and the Hamiltonian formulation of QCD in physical space-time quantized on the light front. 
The light-front wavefunctions of bound states in QCD are relativistic and frame-independent
generalizations of the familiar Schr\"odinger wavefunctions of
atomic physics, but they are determined at fixed light-cone time
$\tau  = t +z/c$---the ``front form" advocated by Dirac~\cite{Dirac:1949cp}---rather than
at fixed ordinary time $t$.  
The  light-front wavefunctions  of a
hadron are independent of the momentum of the hadron, and they are
thus boost invariant; Wigner transformations and Melosh rotations
are not required. The light-front formalism for gauge theories in
light-cone gauge is particularly useful in that there are no ghosts
and one has a direct physical interpretation of  orbital angular
momentum.

An important feature of light-front
quantization is the fact that it provides exact formulas to write
down  matrix elements as a sum of bilinear forms, which can be mapped
into their AdS/CFT counterparts in the semi-classical approximation.
One can thus obtain not only an
accurate description of the hadron spectrum for light quarks, but also
a remarkably simple but realistic model of the valence wavefunctions
of mesons, baryons, and glueballs. 
In terms of light front coordinates $x^\pm =
x^0 \pm x^3$ the AdS metric is
\begin{equation} \label{eq:AdSzLF}
ds^2 = \frac{R^2}{z^2} \left( dx^+ dx^- - d \mbf{x}_\perp^2 - dz^2
\right).
\end{equation}
At fixed light-front time $x^+=0$, the metric depends only on the transverse
$ \mbf{x}_\perp$ and the holographic variable $z$.
Thus we can find an exact correspondence between the
fifth-dimensional coordinate of anti-de Sitter space $z$ and a
specific impact variable $\zeta$ in the light-front formalism.  The
new variable $\zeta$
measures the separation of the constituents within the hadron in
ordinary space-time.  The amplitude $\Phi(z)$ describing  the
hadronic state in $\rm{AdS}_5$ can then be precisely mapped to the
light-front wavefunctions $\psi_{n/h}$ of hadrons in physical
space-time~\cite{Brodsky:2006uqa}, thus providing a relativistic
description of hadrons in QCD at the amplitude level. 
This connection allows one to compute the analytic form~\cite{Brodsky:2006uqa} of the light-front wavefunctions of mesons and
baryons.  AdS/CFT also provides a non-perturbative derivation of dimensional counting rules for the power-law fall-off of form factors and
exclusive scattering amplitudes at large momentum transfer.
The AdS/CFT approach thus leads to a model of hadrons which has both confinement at large distances and the conformal scaling
properties which reproduce dimensional counting rules for hard exclusive reactions.

\section{Gauge/Gravity Semiclassical Correspondence}

The formal statement of the duality between a gravity theory on $(d+1)$-dimensional 
Anti-de Sitter AdS$_{d+1}$ space and the strong coupling limit of a conformal
field theory (CFT)  on the $d$-dimensional asymptotic boundary of AdS$_{d+1}$
at $z=0$ is expressed in terms of the $d+1$ partition function for
a field $\Phi(x,z)$ propagating in the bulk
\begin{equation}
Z_{grav}[\Phi(x,z)] = e^{i S_{eff}[\Phi]} =
\int \mathcal{D}[\Phi] e^{ i S[\Phi]},
\label{eq:Zgrav}
\end{equation}
where $S_{eff}$ is the effective action of the AdS$_{d+1}$ theory,
and the $d$-dimensional generating functional 
of the conformal field theory in presence of an external source $\Phi_0(x)$,
\begin{equation} 
  Z_{CFT}[\Phi_0(x)] =  e^{ i W_{CFT}[\Phi_0]} \\
  = \left< \exp\left(i \int d^dx  \Phi_0(x) \mathcal{O}(x)\right) \right>.
\label{eq:ZCFT}
\end{equation}
The functional $W_{CFT}$ is the generator of connected 
Green's functions of the boundary theory and $\mathcal{O}(x)$ is a QCD local
interpolating operator.
The precise relation of the gravity theory on AdS space
to the conformal field theory at its boundary 
is~\cite{Gubser:1998bc, Witten:1998qj}
\begin{equation}
 Z_{grav}\big[\Phi(x,z) \vert_{z = 0} = \Phi_0(x) \big]
 = Z_{CFT}\left[\Phi_0\right],
\label{eq:grav-CFT}
\end{equation}
where the partition function (\ref{eq:Zgrav}) on AdS$_{d+1}$ is integrated 
over all possible configurations
$\Phi$ in the bulk which approach its boundary value $\Phi_0$.
If we neglect the contributions from the non-classical configurations 
to the gravity partition function, then the generator $W_{CFT}$ of
connected Green's functions of the four-dimensional gauge theory
(\ref{eq:ZCFT}) is precisely equal to the classical (on-shell) gravity action
(\ref{eq:Zgrav})
\begin{equation}
W_{CFT}\left[\phi_0\right] =
S_{eff}\big[\Phi(x,z)\vert_{z=0} = \Phi_0(x)\big]_{\rm on-shell},
\end{equation}
evaluated in terms of the classical solution to the bulk equation of motion.
This defines  the semiclassical approximation to the conformal field theory.
In the limit $z \to 0$, the independent solutions behave as
\begin{equation} \label{eq:Phiz0}
\Phi(z,x) \to z^\Delta \,\Phi_+(x) + z^{d - \Delta} \,\Phi_-(x),
\end{equation}
where $\Delta$ is the conformal dimension.
The non-normalizable solution $\Phi_-$  is the boundary value of the bulk
field $\Phi$ which couples to a QCD gauge invariant operator 
$\mathcal{O}$ in the $z \to 0$ asymptotic boundary, thus $\Phi_- = \Phi_0$.
The normalizable solution $\Phi_+(x)$ is the response
function and corresponds to the physical states~\cite{Balasubramanian:1998sn}. 
The  interpolating
operators $\mathcal{O}$ of the boundary conformal theory are constructed
from local gauge-invariant products of quark and gluon fields and
their covariant derivatives, taken at the same point in
four-dimensional space-time in the $x^2 \to 0$ limit. 
Their conformal twist-dimensions  are matched to the  scaling behavior of
the AdS fields in the limit $z \to 0$ and are thus encoded into
the propagation of the modes inside AdS space. 

\subsection{AdS Wave Equations}

AdS coordinates are the Minkowski coordinates
$x^\mu$ and $z$, the holographic coordinate, which we label 
$x^\ell = (x^\mu, z)$. The metric of the full 
space-time is $ds^2 = g_{\ell m} dx^\ell dx^m$,
where $g_{\ell m} = \left(R^2/z^2\right) \eta_{\ell m}$,
and $\eta_{\ell m}$ has diagonal components
$(1, -1, \cdots, -1)$.
Unless stated otherwise, 
5-dimensional fields are represented by capital letters such as
$\Phi$ and $\Psi$. 
Holographic fields in 4-dimensional Minkowski space are represented by
$\phi$ and $\psi$ and constituent quark and gluon fields by $q$ and
$G$. We begin by writing the action for scalar  modes on AdS$_{d+1}$.
We consider a quadratic action of a free field propagating in the AdS
background
\begin{equation} 
S[\Phi] = \frac{1}{2} \int d^{d+1} x  \sqrt{g} \,
\left[ g^{\ell m} \partial_\ell \Phi\partial_m \Phi
- \mu^2 \Phi^2 \right],
\label{eq:SPhi}
\end{equation} 
where $\sqrt{g} \to (R/z)^{d+1}$ in the conformal limit and $\mu$ is a fifth dimensional mass.
Taking the variation of (\ref{eq:SPhi}) we find the equation of motion
\begin{equation} 
\frac{1}{\sqrt g} \frac{\partial}{\partial x^\ell} 
\bigl(\sqrt g~g^{\ell m} \frac{\partial}{\partial x^m} \Phi \bigr) 
+ \mu^2 \Phi = 0.
\label{eq:EoMPhi}
\end{equation}
Integrating by parts and using the equation of motion, the bulk
contribution to the action vanishes, and one is left with a non-vanishing surface
term in the ultraviolet boundary
\begin{equation} 
S = \frac{ R^{d-1}}{2} \lim_{z \to 0} 
\int d^dx \, \frac{1}{z^{d-1}} \, \Phi \partial_z \Phi, 
\label{eq:SUV}
\end{equation}
which can be identified with the boundary functional $W_{CFT}$. 
Substituting the leading dependence (\ref{eq:Phiz0}) of $\Phi$ near $z =0$  in the ultraviolet
surface action (\ref{eq:SUV}) and using the functional relation
$\delta W_{CFT} /\delta \Phi_0= \delta S_{\rm eff}/\delta\Phi_0$,
it follows that
$\Phi_+(x)$ is related to the expectation values of $\cal O$
in the presence of the source $\Phi_0$~\cite{Balasubramanian:1998sn}:
$ \left\langle 0 \vert {\cal O}(x) \vert 0 \right\rangle_{\Phi_0}
\sim \Phi_+(x).$
The exact relation depends on the normalization of the fields 
used~\cite{Klebanov:1999tb}. The field $\Phi_+$ thus acts as a
classical field, and it is the boundary limit of the normalizable string
solution which propagates in the bulk.

Factoring out the dependence
of the hadronic modes along the Poincar\'e coordinates $x^\mu$, 
$\Phi_P(x,z) = e^{- iP \cdot x} \Phi(z)$ in (\ref{eq:EoMPhi}), 
we find the effective AdS wave
equation for the scalar string mode $\Phi(z)$
\begin{equation} 
\label{eq:AdSPhi}
\left[z^2 \partial_z^2 - (d-1) z\,\partial_z + z^2 \mathcal{M}^2 
- (\mu R)^2 \right] \Phi(z) = 0.
\end{equation} 
The eigenvalues of (\ref{eq:AdSPhi}) are the hadronic invariant mass states 
$P_\mu P^\mu = \mathcal{M}^2$ and the fifth-dimensional mass is related to the
conformal dimension $(\mu R)^2 = \Delta(\Delta - 4)$. Stable solutions satisfy the condition
$(\mu R)^2 \ge - d^2/4$,
according to the Breitenlohner-Freedman
bound~\cite{Breitenlohner:1982jf}. 

Higher spin-$S$ bosonic modes in AdS are described by a set of $S\!+\!1$ coupled 
differential equations~\cite{l'Yi:1998eu}.
Each hadronic state of integer spin $S$, $S \le 2$, is dual to a normalizable string mode
$\Phi_P(x,z)^S _{\mu_1 \mu_2 \cdots \mu_S} =   \epsilon_{\mu_1 \mu_2 \cdots \mu_S}
e^{- i P \cdot x} \, \Phi_S(z)$,
with four-momentum $P_\mu$ and spin polarization
indices along the 3+1 physical coordinates. For string modes with all the polarization indices along the Poncar\'e coordinates,
the coupled differential wave equations for a  spin-$S$ bosonic mode reduce to the  homogeneous equation~\cite{l'Yi:1998eu}  
 \begin{equation} \label{eq:eomPhiSz}
 \left[ z^2 \partial_z^2 - (d\! -\! 1 \!- \!2 S) z \, \partial_z + z^2 \mathcal{M}^2
 \!  -  (\mu R)^2 \right] \!  \Phi_S(z)  = 0,
 \end{equation}
with  $(\mu R)^2 = (\Delta-S)(\Delta-d+S)$.  We expect to avoid large anomalous dimensions
associated with  $S > 2$ since modes with $S \le 2$ do not couple to stringy excitations.

\section{The Holographic Light-Front Hamiltonian and Schr\"odinger Equation}

We shall show in Sect. 5 how the string amplitude $\Phi(z)$ can be mapped
to the light-front wave functions of hadrons in physical space-time~\cite{Brodsky:2006uqa}.
In fact, we find an exact correspondence between the holographic variable $z$ and an
impact variable $\zeta$ which measures the transverse separation of the constituents within
a hadron, we can identify $\zeta = z$. The mapping of $z$ from AdS space to $\zeta$ in the LF space  allows the equations of motion in AdS space to be recast in the form of  a
light-front Hamiltonian equation~\cite{Brodsky:1997de}
\begin{equation}
H_{LF} \ket{\phi} = \mathcal{M}^2 \ket{\phi}, \label{eq:HLC}
\end{equation}
a remarkable result which maps AdS/CFT
solutions to light-front equations in physical 3+1
space-time.  By substituting $\phi(\zeta) =
\zeta^{-3/2} \Phi(\zeta)$, in the AdS scalar wave
equation (\ref{eq:AdSPhi}) for $d = 4$,
we find an effective Schr\"odinger equation as a function of the
weighted impact variable $\zeta$
\begin{equation} \label{eq:Scheq}
\left[-\frac{d^2}{d \zeta^2} + V(\zeta) \right] \phi(\zeta) =
\mathcal{M}^2 \phi(\zeta),
\end{equation}
with the conformal potential $V(\zeta) \to - (1-4 L^2)/4\zeta^2$,
an effective two-particle light-front radial equation for mesons~\cite{Brodsky:2007pt, Brodsky:2006uqa}.
Its eigenmodes determine the hadronic mass spectrum. 
We have written above $(\mu R)^2 = - 4 + L^2$. 
The holographic hadronic light-front
wave functions $\phi(\zeta) = \langle \zeta \vert \phi \rangle$ are
normalized according to
\begin{equation}
\langle \phi \vert \phi \rangle = \int d\zeta \, \vert \langle
\zeta \vert \phi \rangle \vert^2 = 1,
\end{equation}
and represent the probability amplitude to find $n$-partons at
transverse impact separation $\zeta = z$.   Its
eigenvalues are set by the boundary conditions at 
$\phi(z =1/\Lambda_{\rm QCD}) = 0$ and are given in terms of the roots of
Bessel functions: $\mathcal{M}_{L,k} = \beta_{L,k} \Lambda_{\rm
QCD}$. The normalizable modes are
\begin{equation}
\phi_{L,k}( \zeta) =   \frac{ \sqrt{2} \Lambda_{\rm QCD}}{J_{1+L}(\beta_{L,k})}
 \sqrt{\zeta} J_L \! \left(\zeta \beta_{L,k} \Lambda_{\rm QCD}\right)
 \theta\big(\zeta \le
\Lambda^{-1}_{\rm QCD}\big).
\end{equation}

The lowest stable state $L = 0$ is determined by the
Breitenlohner-Freedman bound~\cite{Breitenlohner:1982jf}. 
Higher excitations are matched to the small $z$ asymptotic behavior of each string mode
to the corresponding
conformal dimension of the boundary operators
of each hadronic state. The effective wave equation
(\ref{eq:Scheq}) is a relativistic light-front equation defined at
$x^+ = 0$. The AdS metric $ds^2$ (\ref{eq:AdSzLF})  is invariant if
$\mbf{x}_\perp^2 \to \lambda^2 \mbf{x}_\perp^2$ and $z \to \lambda
z$ at equal light-front time  $x^+ = 0$. The Casimir operator for the rotation
group $SO(2)$ in the transverse light-front plane is $L^2$. This
shows the natural holographic connection to the light front.

For higher spin bosonic modes  we can also recast the wave equation AdS (\ref{eq:eomPhiSz}) into its light-front form (\ref{eq:HLC}). Using the substitution  
$\phi_S(\zeta) = \zeta^{-3/2+S} \Phi_S(\zeta)$, $\zeta = z$, we find a LF Schr\"odinger equation
identical to (\ref{eq:Scheq}) with $\phi  \to \phi_S$,
provided that $(\mu R)^2 = - (2-S)^2 +\nu^2$.
Stable solutions satisfy a generalized Breitenlohner-Freedman bound
$(\mu R)^2 \ge -  (d - 2 S)^2/4$,
and thus the lowest stable state has scaling dimensions $\Delta = 2$,
 independent of $S$.
The fundamental LF equation of AdS/CFT has the appearance of a
Schr\"odinger  equation, but it is relativistic, covariant, and analytically tractable.

The pseudoscalar meson interpolating operator
$\mathcal{O}_{2+L}= \bar q \gamma_5 D_{\{\ell_1} \cdots D_{\ell_m\}} q$, 
written in terms of the symmetrized product of covariant
derivatives $D$ with total internal space-time orbital
momentum $L = \sum_{i=1}^m \ell_i$, is a twist-two, dimension $3 + L$ operator
with scaling behavior determined by its twist-dimension $ 2 + L$. Likewise
the vector-meson operator
$\mathcal{O}_{2+L}^\mu = \bar q \gamma^\mu D_{\{\ell_1} \cdots D_{\ell_m\}} q$
has scaling dimension $2 + L$.  The scaling behavior of the scalar and vector AdS modes is precisely the scaling required to match the scaling dimension of the local pseudoscalar and vector-meson interpolating operators. The light meson spectrum is compared in Figure \ref{fig:MesonSpec}
with the experimental values.
\begin{figure}[h]
\centering
\includegraphics[angle=0,width=9.6cm]{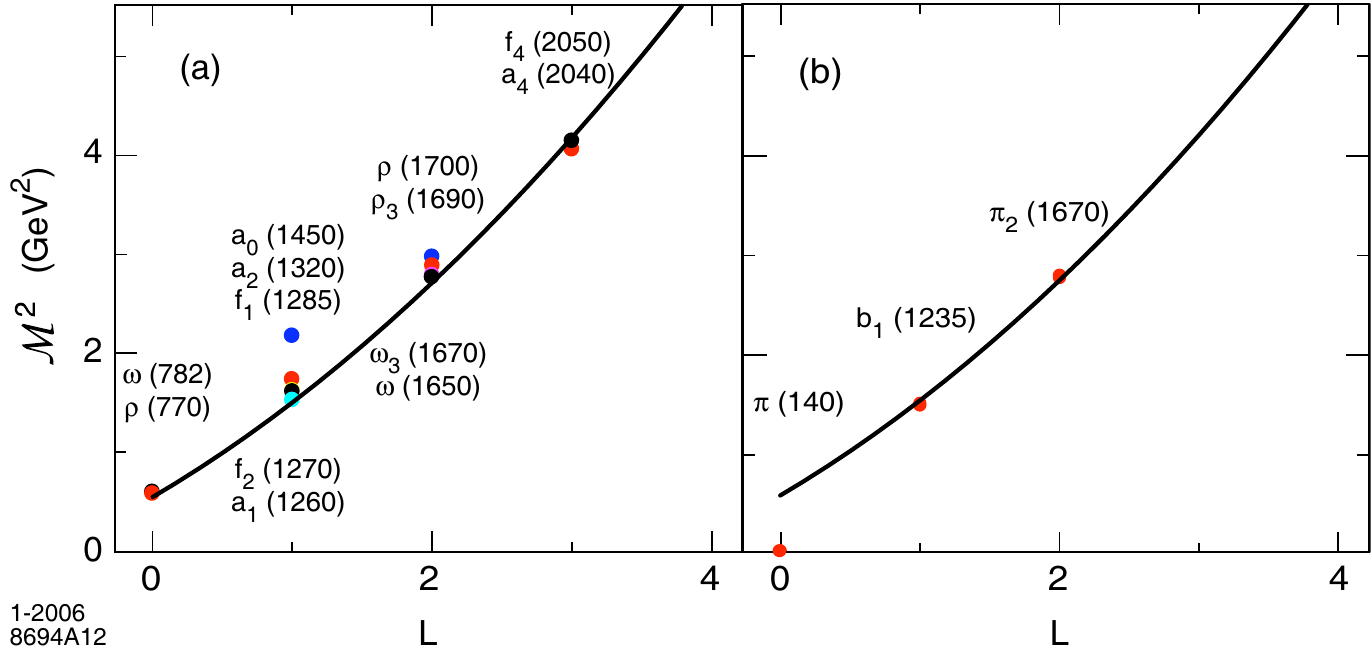}
\caption{Light meson orbital states for $\Lambda_{\rm QCD}$ = 0.32 GeV:
(a) vector mesons and (b) pseudoscalar mesons. The data are from \cite{Yao:2006px}.}
\label{fig:MesonSpec}
\end{figure}

\subsection{Integrability of AdS/CFT Equations}

The integrability methods of \cite{Infeld:1941} find a remarkable application in the AdS/CFT
correspondence.   Integrability follows
if the equations describing a physical model can be factorized in
terms of linear operators. These ladder operators
generate all the eigenfunctions once the lowest mass eigenfunction is known.
In holographic QCD, the conformally invariant 3 + 1 light-front differential equations can be expressed in terms of ladder operators and their solutions can then be expressed in terms of analytical functions.
In the conformal limit the ladder algebra for bosonic ($B$) or fermionic ($F$) modes is given in terms of
the operator ($\Gamma^B =1, ~~ \Gamma^F = \gamma_5$)
\begin{equation} \label{eq:Pi}
\Pi_\nu^{B,F}(\zeta) = -i \left( \frac{d}{d \zeta} - \frac{\nu + \half}{\zeta} \, \Gamma^{B,F} \!\right) ,
\end{equation}
and its adjoint
\begin{equation}
\Pi^{B,F}_\nu(\zeta)^\dagger = -i \left(\frac{d}{d \zeta} + \frac{\nu + \half}{\zeta}
\, \Gamma^{B,F} \!\right) ,
\end{equation}
with commutation relations
\begin{equation}
\left[\Pi_\nu^{B,F}(\zeta),\Pi_\nu^{B,F}(\zeta)^\dagger\right]
=  \frac{2 \nu+1}{\zeta^2} \, \Gamma^{B,F} .
\end{equation}
For bosonic modes  the Hamiltonian is written as a bilinear form:
$H^{B,F}_{LC} = {\Pi_\nu^{B,F}}^\dagger \Pi_\nu^{B,F}$. For $\nu^2 \ge 0$ the Hamiltonian is positive definite
\begin{equation} 
\langle \phi \left\vert H_{LC}^\nu \right\vert \phi \rangle
= \int d\zeta \, \vert \Pi_\nu \phi(z) \vert^2 \ge 0,
\end{equation}
and its eigenvalues are positive:
$\mathcal{M}^2 \ge 0$.  For $\nu^2 < 0$
the Hamiltonian is not bounded from
below.  The critical value of the potential corresponds to $\nu = 0$ with 
potential $V_{crit}(\zeta) = 1/4 \zeta^2$. LF quantum-mechanical stability conditions are thus equivalent
to the stability conditions which follows from the
Breitenlohner-Freedman stability bound~\cite{Breitenlohner:1982jf}.
Higher orbital states
are constructed from the $L$-th application of the raising operator $a^\dagger = - i \Pi^B$ on the
ground state  $\vert L \rangle  \sim (a^\dagger)^L \vert 0 \rangle$.
In the $\zeta$ light-front coordinate representation \vspace{-7pt}
 \begin{equation}
 \langle \zeta \vert L \rangle 
 \sim  \sqrt{\zeta} \, (-\zeta)^L \left(\frac{1}{\zeta} \frac{d}{d \zeta}\right)^L
  J_0(\zeta \mathcal{M}) \sim
   \sqrt{\zeta} J_L\left(\zeta \mathcal{M}\right).
  \end{equation}
In the fermionic case the eigenmodes
also satisfy a first order LF Dirac equation as will be shown in Sect. 4.

\subsection{Soft-Wall Holographic Model}

The predicted mass spectrum in the truncated space hard-wall (HW) model is
linear $M \propto L + 2 \, n $ at high orbital
angular momentum $L$, in contrast to the quadratic dependence $M^2 \propto L + n$ in the usual
Regge parameterization. 
It has been shown recently that by  choosing a specific profile for a non-constant dilaton, the usual Regge  dependence  can be obtained~\cite{Karch:2006pv}.  
This procedure 
retains conformal AdS metrics (\ref{eq:AdSz}) while introducing a smooth cutoff  which depends on the profile of a dilaton background field $\varphi$ 
\begin{equation}
S = \int \! d^4x \, dz  \,\sqrt{g} \,e^{- \varphi(z) } \mathcal{L},
\end{equation}
where $\varphi$ is a function of the holographic coordinate $z$ which vanishes in the ultraviolet 
limit $z \to 0$. The IR hard-wall or truncated space holographic model
corresponds to a constant dilaton field $\varphi(z) \! = \! \varphi_0$ in the confining region,  
$z \leq 1/\Lambda_{\rm QCD}$, and to very large values elsewhere: $\varphi(z) \to \infty$ for 
$z >  1/\Lambda_{\rm QCD}$. The introduction of a soft cutoff avoids the
ambiguities in the choice of boundary conditions at the infrared wall.  A convenient choice~\cite{Karch:2006pv} for the background field with
usual Regge behavior is $\varphi(z) = \kappa^2 z^2$.
The resulting wave equations are equivalent to the radial equation of a two-dimensional
oscillator, previously found in the context  of mode propagation on 
AdS$_5 \times S^5$, in the light-cone formulation of Type II supergravity~\cite{Metsaev:1999kb}.
Also,  equivalent results follow from
the introduction of a gaussian warp factor in the AdS metric
for the particular case of  massless vector modes propagating in the distorted 
metric~\cite{Andreev:2006vy}. A different approach to the soft-wall (SW) consists in the non-conformal
extension of the algebraic expressions found
in the previous section to obtain directly the corresponding holographic LF wave equations.  This method is particularly useful to extend the non-conformal results to the fermionic sector
where the corresponding linear wave equations become exactly solvable. The extended generators
are given 
in terms of the matrix-valued operator $\Pi$ and its adjoint $\Pi^\dagger$ 
($\Gamma^B =1, ~~ \Gamma^F = \gamma_5$)
\begin{eqnarray} \label{A}
\Pi_\nu^{B,F}(\zeta) &\!=\!& -i\left( \frac{d}{d \zeta} 
- \frac{\nu + \half}{\zeta} \Gamma^{B,F} - \kappa^2 \zeta \,\Gamma^{B,F} \right),\\
\Pi^{B,F}_\nu(\zeta)^\dagger &\!=\!& -i\left(\frac{d}{d \zeta} 
+ \frac{\nu + \half}{\zeta} \Gamma^{B,F} + \kappa^2 \zeta \, \Gamma^{B,F}\right),
\end{eqnarray}
with commutation relations
\begin{equation}
\left[\Pi^{B,F}_\nu(\zeta),\Pi^{B,F}_\nu(\zeta)^\dagger\right] =  
\left(\frac{2\nu+1}{\zeta^2} - 2 \kappa^2\right) \!\Gamma^{B,F}.
\end{equation}
An account of the extended algebraic holographic model and a possible supersymmetric
connection between the bosonic and fermionic operators used in the holographic
construction will be described elsewhere.

\section{Baryonic Spectra in AdS/QCD}

The holographic model
based on truncated AdS space can be used to obtain the hadronic
spectrum of light quark $q \bar q, qqq$ and $gg$ bound states.  Specific
hadrons are identified by the correspondence of the AdS amplitude  with
the twist dimension of the interpolating operator for the hadron's valence
Fock state, including its orbital angular momentum excitations.   Bosonic modes with conformal
dimension $2+L$ are dual to the interpolating operator $\mathcal{O}_{\tau + L}$ with $\tau = 2$.
For fermionic modes $\tau = 3$.

As an example, we will outline here the analysis of the baryon spectrum in AdS/CFT.
The action for massive fermionic  modes on AdS$_{d+1}$ is
 \begin{equation} 
 S[\bar \Psi, \Psi] = \int \! d^{d+1} x   \, \sqrt{g} \, 
\bar \Psi \left(i \Gamma^\ell D_\ell - \mu\right) \Psi,
 \end{equation} 
with the equation of motion
 \begin{equation} 
 \left[i\left( z \eta^{\ell m} \Gamma_\ell \partial_m + \frac{d}{2} \Gamma_z \right)
 + \mu R  \right] \Psi(x^\ell) = 0 .
 \end{equation}
Upon the substitution
~$\Psi(x,z) = e^{-i P \cdot x} z^2 \psi(z)$, ~$z \to \zeta$,
we find  the light-front Dirac equation
\begin{equation} \label{eq:LFDE}
\left(\alpha \, \Pi^F \! (\zeta) - \mathcal{M} \right)  \psi(\zeta) = 0,
\end{equation}
where the generator $\Pi^F$ is given by (\ref{eq:Pi}) and
$
 i \alpha =
  \begin{pmatrix}
  0& I\\
- I& 0
  \end{pmatrix}
$ in the Weyl representation.
The solution is\begin{equation} \label{eq:DiracLF}
\psi(\zeta) =  C \sqrt{\zeta}
\left[J_{L+1}\left(\zeta \mathcal{M} \right) \, u_+
+ J_{L+2}\left(z \mathcal{M}\right) \, u_- \right],
\end{equation}
with $\gamma_5 u_\pm = u_\pm$.
A discrete  four-dimensional spectrum follows when we impose the boundary condition
$\psi_\pm(\zeta=1/\Lambda_{\rm QCD}) = 0$:
$\mathcal{M}_{\alpha, k}^+ = \beta_{\alpha,k} \Lambda_{\rm QCD}, ~~
\mathcal{M}_{\alpha, k}^- = \beta_{\alpha + 1,k} \Lambda_{\rm QCD}$,
with a scale-independent mass ratio~\cite{deTeramond:2005su}.

Figure \ref{fig:BaryonSpec}(a) shows the predicted orbital spectrum of the
nucleon states and Fig. \ref{fig:BaryonSpec}(b) the $\Delta$ orbital
resonances. The spin-3/2 trajectories are determined from the corresponding Rarita-Schwinger  equation.  The solution of the spin-3/2 for polarization along Minkowski coordinates, $\psi_\mu$, is 
identical to the spin-1/2 solution.
The data for the baryon spectra are from~\cite{Yao:2006px}.
The internal parity of states is determined from the SU(6)
spin-flavor symmetry.
\begin{figure}[ht]
\centering
\includegraphics[angle=0,width=11.6cm]{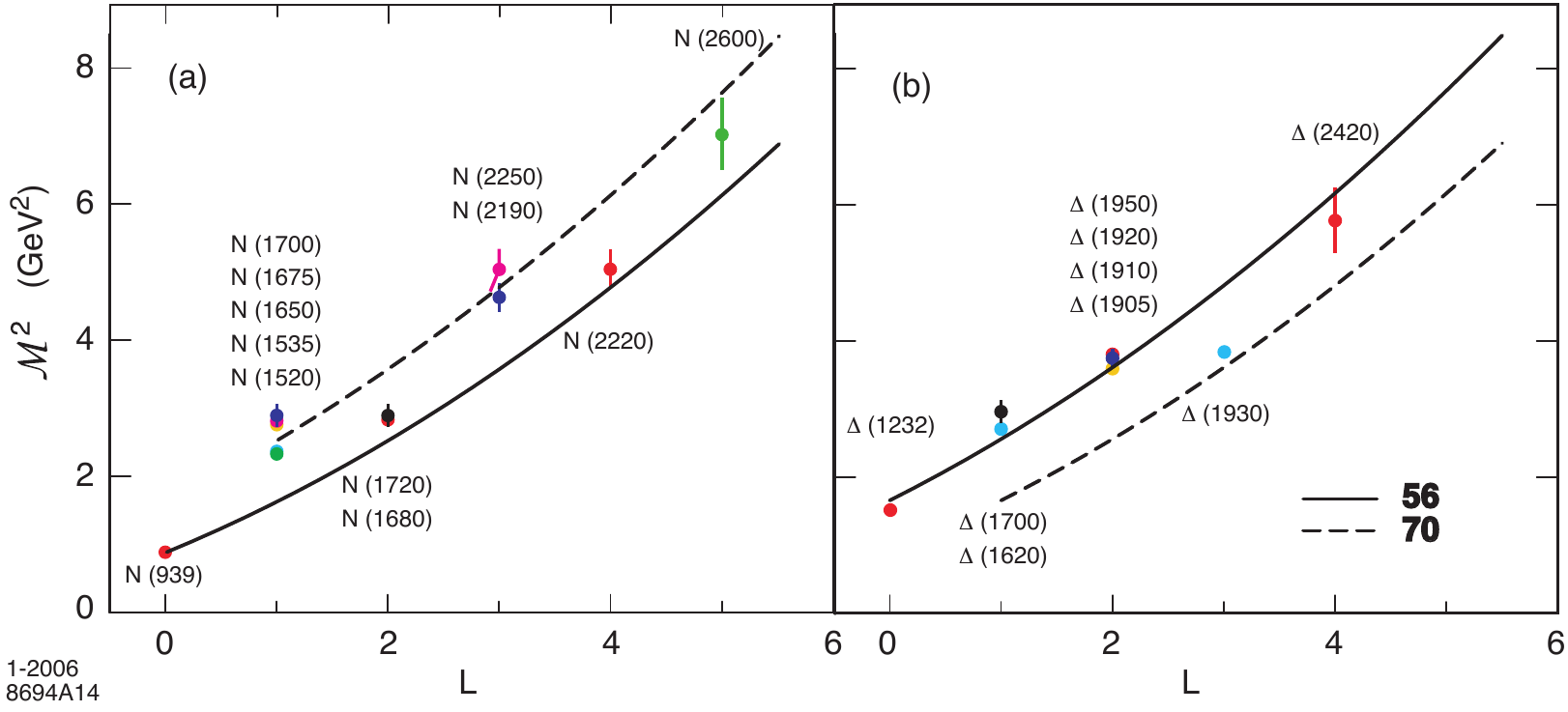}
\caption{Predictions for the light baryon orbital spectrum for
$\Lambda_{QCD}$ = 0.25 GeV. The  $\bf 56$ trajectory corresponds to
$L$ even  $P=+$, and the $\bf 70$ to $L$ odd  $P=-$ states.}
\label{fig:BaryonSpec}
\end{figure}
Since only
one parameter, the QCD mass scale $\Lambda_{\rm QCD}$  is introduced, the
agreement with the pattern of physical states is remarkable. In
particular, the ratio of $\Delta$ to nucleon trajectories is
determined by the ratio of zeros of Bessel functions.  

We can solve the LF Dirac equation (\ref{eq:LFDE}) with the non-conformal extended
generator $\Pi^F$ given by (\ref{A}). The solutions to the Dirac equation are
\begin{eqnarray}
\psi_+(\zeta) &\sim& z^{\frac{1}{2} + \nu} e^{-\kappa^2 \zeta^2/2}
  L_n^\nu(\kappa^2 \zeta^2) ,\\
\psi_-(\zeta) &\sim&  z^{\frac{3}{2} + \nu} e^{-\kappa^2 \zeta^2/2}
 L_n^{\nu+1}(\kappa^2 \zeta^2).
\end{eqnarray}
with eigenvalues
$\mathcal{M}^2 = 4 \kappa^2( n + \nu + 1)$.  
Comparing with usual Dirac equation in AdS space  we find 
\begin{equation} 
\left[i\left( z \eta^{\ell m} \Gamma_\ell \partial_m + \frac{d}{2} \Gamma_z \right)
 + \mu R + V(z) \right] \Psi(x^\ell) = 0 .
\end{equation}
with $V(z) = \kappa^2 z.$
Thus for fermions the ``soft-wall" corresponds to fermion modes propagating in AdS conformal metrics in presence of a  linear confining potential.

\section{Hadronic Form Factors in AdS/QCD}

The AdS/QCD correspondence is particularly relevant for the description
of hadronic form factors,  since it incorporates the connection between the twist of the hadron to the fall-off of its current matrix elements, as well as  essential aspects of vector meson dominance.
It also provides a convenient framework for analytically continuing the space-like results to the time-like region. Recent applications to the electromagnetic~\cite{deTeramond:2006xb, Radyushkin:2006iz, Grigoryan:2007vg, Grigoryan:2007my, Brodsky:2007hb, Brodsky:2007vk, 
Huang:2007uu, Kwee:2007dd}
and gravitational~\cite{Abidin:2008ku} form factors  of hadrons
have followed from the original work described
in~\cite{Polchinski:2002jw,Hong:2004sa}.

\subsection{Meson Form Factors}

In AdS/CFT, the hadronic matrix element for the electromagnetic current 
has the form of a convolution of the string modes for the initial and final
hadrons with the external electromagnetic source which propagates inside AdS.
We discuss first the truncated space or hard wall~\cite{Polchinski:2001tt} 
holographic model, where  quark and gluons as well as the external electromagnetic 
current propagate freely into the AdS interior according to
the AdS metric. Assuming minimal coupling the form factor
has the form~\cite{Polchinski:2002jw, Hong:2004sa}
\begin{equation}
i g_5 \int d^4x \, dz \,\sqrt{g}\, A^{\ell}(x,z)
\Phi_{P'}^*(x,z) \overleftrightarrow\partial_\ell \Phi_P(x,z),
\label{eq:M}
\end{equation} 
where $g_5$ is a five-dimensional effective coupling constant and
$\Phi_P(x,z)$ is a normalizable mode representing a hadronic state,
$\Phi_P(x,z) \sim e^{-iP \cdot x} \Phi(z)$, with hadronic invariant mass
given by $P_\mu P^\mu = \mathcal{M}^2$.
We consider the propagation inside AdS space of an electromagnetic
probe polarized along Minkowski coordinates  ($Q^2 = - q^2 > 0$)
$A(x,z)_\mu = \epsilon_\mu e^{-i Q \cdot x} J(Q^2,z),~ A_z = 0$,
where $J(Q^2, z)$  has the value 1 at
zero momentum transfer, since we are normalizing the bulk solutions to the total charge operator,
and as boundary limit the external current 
$A_\mu(x, z \to 0) = \epsilon_\mu e^{-i Q \cdot x}$.
Thus $J(Q^2 = 0, z) = J(Q^2 ,z = 0) = 1$.

The propagation of the external current inside AdS  space is described by the
wave equation 
\begin{equation} \label{eq:AdSJ}
\left[ z^2 \partial_z^2 -  z \, \partial_z - z^2 Q^2 \right]   J(Q^2, z)  = 0,
 \end{equation}
with the solution $J(Q^2, z) = z Q K_1(z Q)$.
Substituting the normalizable mode $\Phi(x,z)$ in (\ref{eq:M}) and extracting an overall delta
function from momentum conservation at the vertex, we find 
the matrix element
$\left\langle P' \left\vert J^\mu(0) \right\vert P \right\rangle = 2
(P + P')^\mu F(Q^2)$, with
\begin{equation} 
F(Q^2)  =  R^3  \! \int \frac{dz}{z^3} \, \Phi(z) J(Q^2, z) \Phi(z).
\label{eq:FFAdS}
\end{equation}
The form factor in AdS is thus 
represented as the overlap 
of the normalizable modes dual to the incoming
and outgoing hadrons, $\Phi_P$ and $\Phi_{P'}$, with the
non-normalizable mode, $J(Q^2, z)$, dual to the external
source~\cite{Polchinski:2002jw}.
Since $K_n(x) \sim \sqrt{\pi/2 x} \, e^{-x}$ for large $x$, it follows
that the external electromagnetic field is suppressed inside the AdS cavity
for large $Q$. At small $z$ the string modes
scale as $\Phi \sim z^\Delta$.
At large enough $Q$, the important contribution to (\ref{eq:FFAdS})
is from the region near $z \sim 1/Q$: 
$F(Q^2) \to \left(1/Q^2\right)^{\Delta - 1}$,
and the ultraviolet point-like behavior~\cite{Polchinski:2001ju} responsible for the power law scaling~\cite{Brodsky:1974vy, Matveev:1973ra} is recovered. This is a remarkable consequence of truncating AdS space since we are describing the coupling of an electromagnetic
current to an extended mode, and instead of soft collision amplitudes characteristic of strings, hard point-like ultraviolet behavior is found~\cite{Polchinski:2001tt}.

The form factor in
AdS space in presence of the dilaton background $\varphi = \kappa^2 z^2$
has the additional term $e^{- \kappa^2 z^2}$  in the metric 
\begin{equation} 
F(Q^2) = R^3 \int \frac{dz}{z^3} \, e^{- \kappa^2 z^2} \Phi(z) J_\kappa(Q^2, z) \Phi(z).
\label{eq:FFAdSSW}
\end{equation}
Since the non-normalizable modes also couple
to the dilaton field, we must study the solutions of the modified wave equation describing  the propagation in AdS space of an electromagnetic probe.
The solution is \cite{Grigoryan:2007my, Brodsky:2007hb}
\begin{equation} \label{eq:Jkappa}
J_\kappa(Q^2, z) = \Gamma\!\left(1+ \frac{Q^2}{4 \kappa^2}\right) 
U\!\left(\frac{Q^2}{4 \kappa^2}, 0 , \kappa^2 z^2\right),
\end{equation}
where $U(a,b,c)$ is the confluent hypergeometric function with the
integral representation
$\Gamma(a)  U(a,b,z) =  \int_0^\infty \! e^{- z t} t^{a-1}
(1+t)^{b-a-1} dt.$ 
In the large $Q^2$ limit, $Q^2 \gg 4 \kappa^2$  we find that
$J_\kappa(Q,z) \to z Q K_1(z Q)$. Thus, for large transverse
momentum the current decouples from the dilaton background.

We can compute the pion form factor from the AdS expressions  (\ref{eq:FFAdS}) and (\ref{eq:FFAdSSW})
for the hadronic string modes $\Phi_\pi$ in the hard-wall (HW)
\begin{equation} \label{eq:PhipiHW}
\Phi_\pi^{HW}(z) = \frac{\sqrt{2} \Lambda_{QCD}}{R^{3/2} J_1(\beta_{0,1}) }
z^2 J_0\left(z \beta_{0,1} \Lambda_{QCD} \right) ,
\end{equation}
and soft-wall (SW) model
\begin{equation} \label{eq:PhipiSWa}
\Phi_\pi^{SW}(z) = \frac{\sqrt{2} \kappa}{R^{3/2}}\, z^2 ,
\end{equation}
respectively.  For the soft wall model the results for form factors can be expressed 
analytically. For integer twist $\tau = n$ the form factor is expressed as a $N-1$ product of poles,
corresponding to the first $n-1$ states along the vector meson trajectory~\cite{Brodsky:2007hb}.
Since the pion mode couples to a twist-two boundary
interpolating operator which creates a  two-component hadronic bound state, the
form factor is given in the SW model by a simple monopole form.  
In Fig. \ref{fig:PionQ2FFSL}, we plot
the product $Q^2 F_\pi(Q^2)$ for the soft and hard-wall holographic models.  
\begin{figure}[h]
\centering
\includegraphics[angle=0,width=7.0cm]{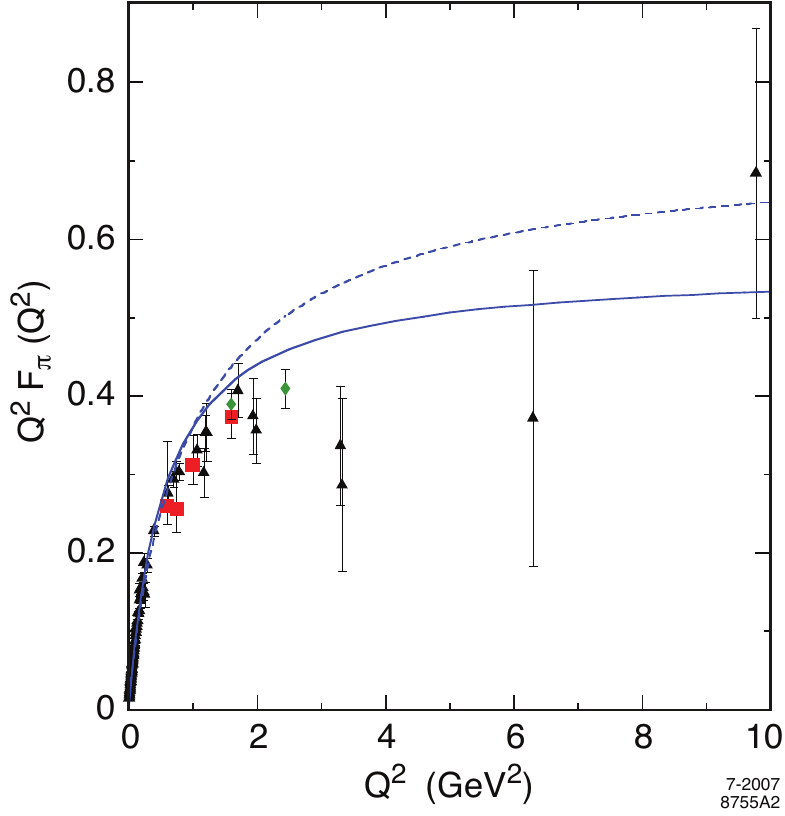}
\caption{Space-like scaling behavior for $Q^2 F_\pi(Q^2)$ as a function of $Q^2 = -q^2$. The continuous line is the prediction of the soft-wall model for  $\kappa = 0.375$ GeV. The dashed line is the prediction of the hard-model for $\Lambda_{QCD} = 0.22$ GeV. The black triangles are from the data compilation of Baldini  {\it et al.}~\cite{Baldini:1998qn},  and the red boxes   and cobalt green diamonds are JLAB   data~\cite{Tadevosyan:2007yd} . }
\label{fig:PionQ2FFSL}
\end{figure}
When the results for the pion form factor are analytically continued
to the time-like region, $q^2 \to -q^2$ we obtain the results shown
in Figure \ref{fig:PionFFLog} for $\log\left(\vert
F_\pi(q^2)\vert\right)$ in the SW model. The monopole form of the SW model exhibits
a pole at the $\rho$ mass and reproduces  well the $\rho$ peak  with
$M_\rho = 4 \kappa^2 = 750$ MeV. In the strongly coupled
semiclassical gauge/gravity limit hadrons have zero widths  and are
stable. The form factor accounts for the scaling behavior in the
space-like region, but it does not give rise to the additional
structure found in the time-like region since the $\rho$ pole
saturates 100\% of  the monopole form.
\begin{figure}[h ]
\centering
\includegraphics[angle=0,width=7.6cm]{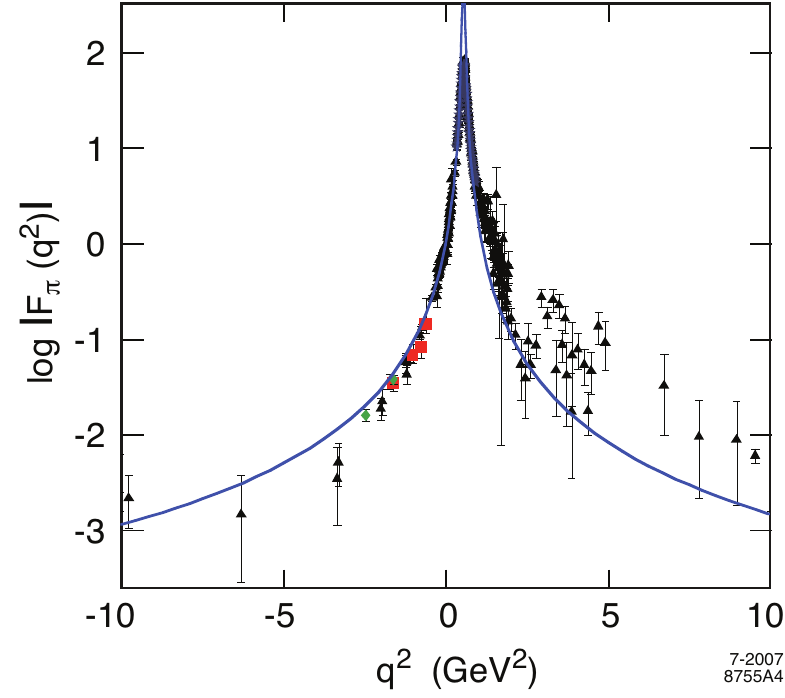}
\caption{Space and time-like behavior of the pion form factor $\log\left(\vert F_\pi(q^2)\vert\right)$ as a function of $q^2$ for $\kappa = 0.375$ GeV in the  soft-wall model. The black (triangle) is from the data compilation of  Baldini  {\it et al.}~\cite{Baldini:1998qn},  and the red (box) and cobalt green diamonds are JLAB data~\cite{Tadevosyan:2007yd}.}
\label{fig:PionFFLog}
\end{figure}

\subsection{The Nucleon Dirac Form Factors}

As an example of a twist $\tau = 3$ fall-off we compute the spin non-flip 
nucleon form factor in the soft wall
model. Consider the spin non-flip form factors
 \begin{eqnarray}
 F_+(Q^2) &\!=\!&  g_+ \, R^4 \int \frac{d z}{z^4}  \, e^{- \kappa^2 z^2} J_\kappa(Q, z) \,
  \vert \Psi_+(z)\vert^2 ,\\
 F_-(Q^2) &\!=\!&  g_- \, R^4\int  \frac{d z}{z^4} \, e^{- \kappa^2 z^2} J_\kappa(Q, z)  \,
  \vert \Psi_-(z)\vert^2 ,
 \end{eqnarray}
 where the effective charges $g_+$ and $g_-$ are determined from the spin-flavor structure of
 the theory.
 We choose the struck quark to have $S^z=+1/2$. The two AdS solutions $\Psi_+$ and $\Psi_-$
correspond to nucleons with total angular momentum $J^z= +1/2$ and $-1/2$.
For the $SU(6)$ spin-flavor symmetry
\begin{eqnarray}
F_1^p(Q^2) &\!=\!&  R^4 \! \int \frac {d z}{z^4}  \, e^{- \kappa^2 z^2} J_\kappa(Q, z) \,
  \vert \Psi_+(\zeta)\vert^2 ,\\
F_1^n(Q^2) &\!=\!& - \frac{1}{3} R^4  \! \int  \frac{d z}{z^4}  \, e^{- \kappa^2 z^2} J_\kappa(Q, z) 
 \left[\vert \Psi_+(z)\vert^2 - \vert\Psi_-(z)\vert^2\right],
 \end{eqnarray}
where $F_1^p(0) = 1$,~ $F_1^n(0) = 0$. The bulk-to-boundary
propagator $J_\kappa(Q,z)$ is the solution (\ref{eq:Jkappa}) of the
AdS wave equation for the external electromagnetic current,  and the
plus and minus components of the twist 3 nucleon mode in the SW model are
\begin{equation} \label{eq:PhipiSW}
\Psi_+(z) \!=\! \frac{\sqrt{2} \kappa^2}{R^{2}}\, z^{7/2},  ~~~~~~
\Psi_-(z) \!=\! \frac{ \kappa^3}{R^{2}}\, z^{9/2}.
\end{equation}
For the SW model the results for $Q^4 F_1^p(Q^2)$ and $Q^4 F_1^n(Q^2)$ follow from the analytic form for the form factors for any $\tau$ given in Appendix D of reference~\cite{Brodsky:2007hb}  and are shown in
Figure \ref{fig:nucleonFF}.
\begin{figure}[ht]
\centering
\includegraphics[angle=0,width=7.55cm]{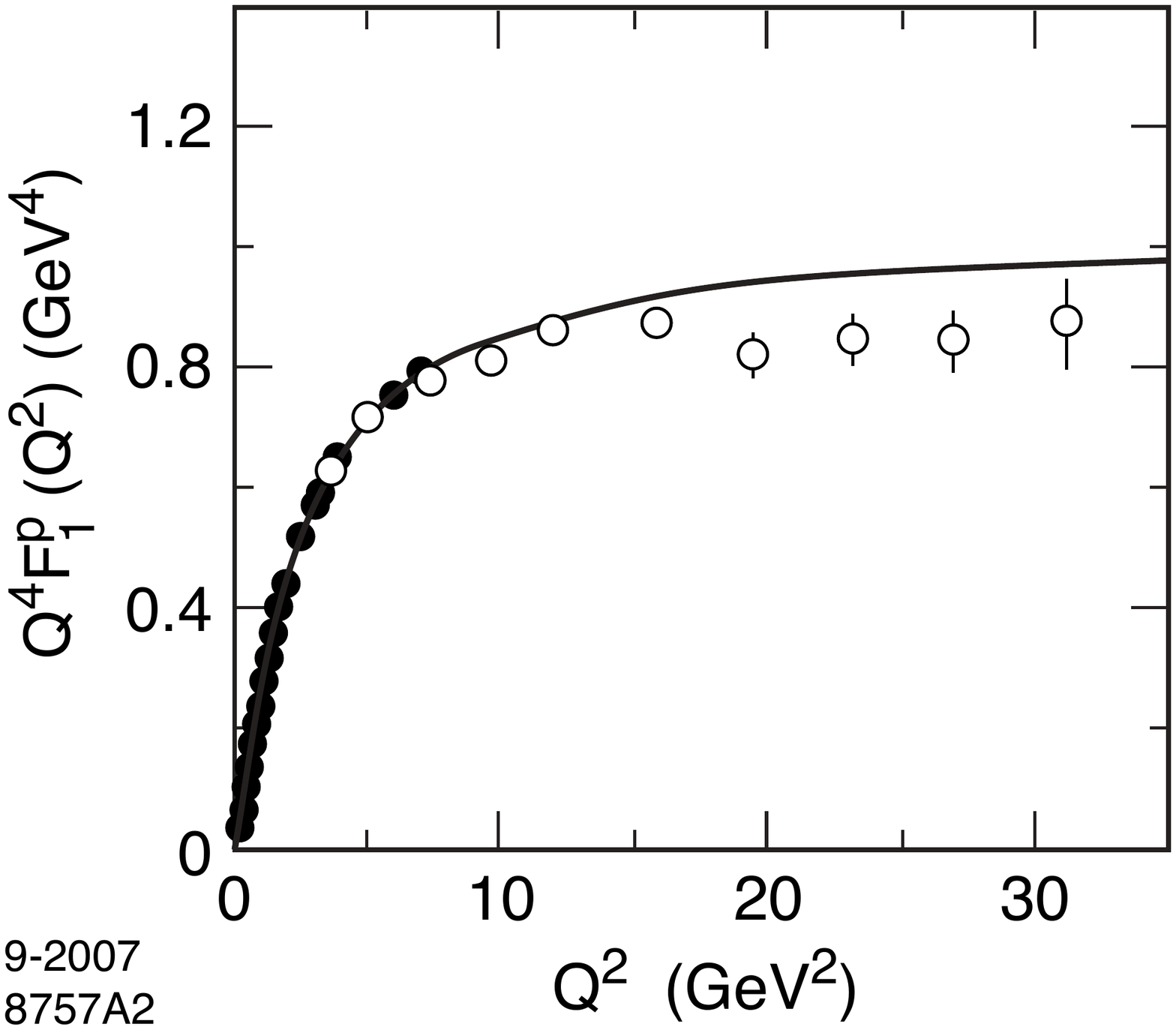}
\includegraphics[angle=0,width=7.5cm]{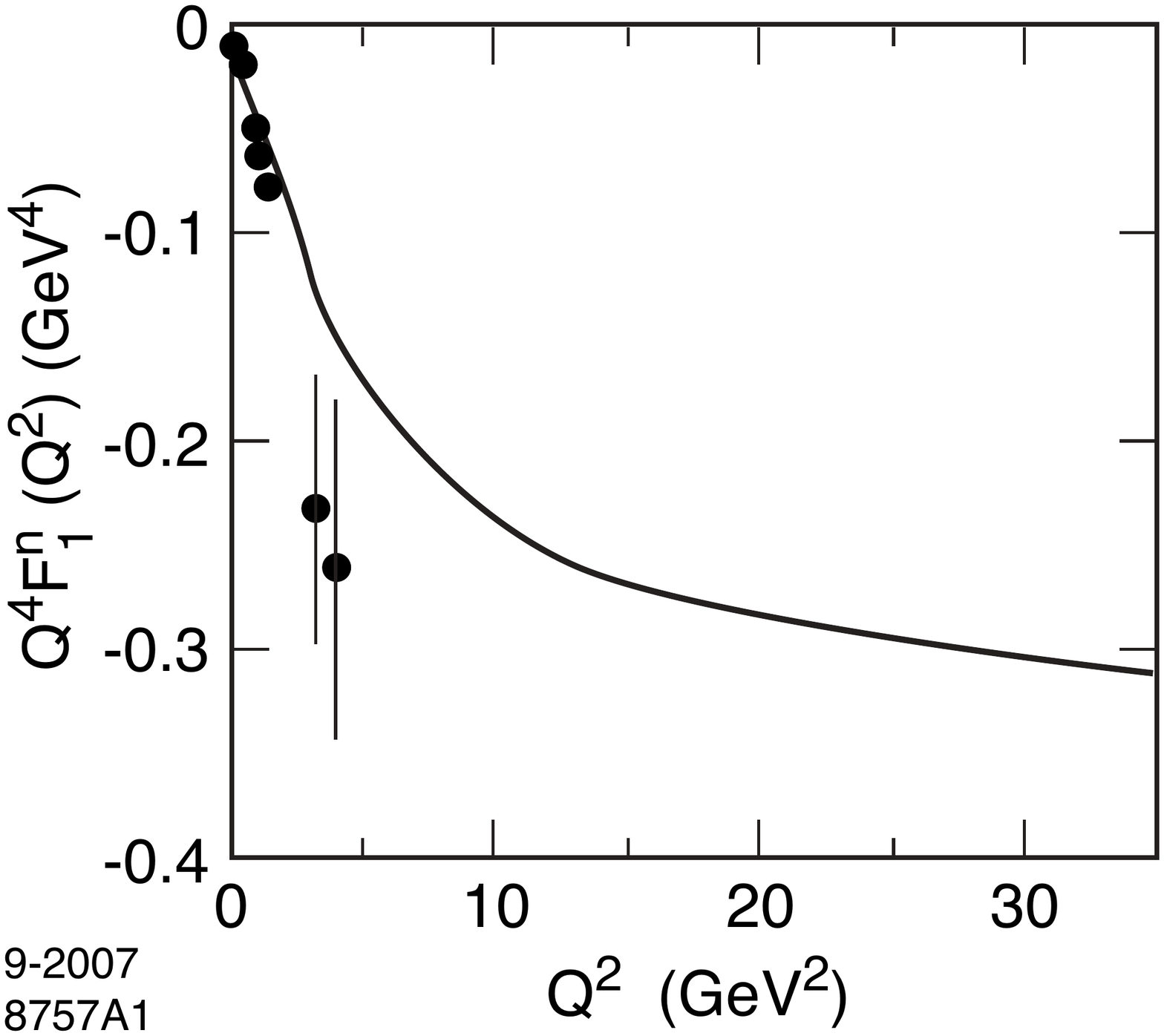}
\caption{Predictions for $Q^4 F_1^p(Q^2)$ and $Q^4 F_1^n(Q^2)$ in the
soft wall model for $\kappa =  0.424$ GeV.  The data compilation is
from Diehl~\cite{Diehl:2005wq}.} \label{fig:nucleonFF}
\end{figure}

\section{The Light-Front Fock Representation}

The light-front expansion of any hadronic system
is constructed by quantizing quantum chromodynamics
at fixed light-cone time \cite{Dirac:1949cp} $\tau = t + z/c$.
In terms of the hadron  four-momentum $P = (P^+, P^-, \mbf{P}_{\!\perp})$,
$P^\pm = P^0 \pm P^3$,
the light-cone Lorentz invariant Hamiltonian for the composite system, 
$H_{LF}^{QCD} = P^-P^+ - \mbf{P}^2_\perp$,  has
eigenvalues given in terms of the eigenmass ${\cal M}$ squared  corresponding 
to the mass spectrum of the color-singlet states in QCD~\cite{Brodsky:1997de}.

The hadron wavefunction is an eigenstate of the total momentum $P^+$
and $\mbf{P}_{\! \perp}$ and the longitudinal spin projection $S_z$,
and is normalized according to
\begin{equation}
\bigl\langle \psi_h(P^+,\mbf{P}_{\! \perp}, S_z) \big\vert 
\psi_h(P'^+,\mbf{P}'_\perp,S_z') \bigr\rangle 
= 2 P^+ (2 \pi)^3 \,\delta_{S_z,S'_z} \,\delta \bigl(P^+ - P'^+ \bigr)
\,\delta^{(2)} \negthinspace \bigl(\mbf{P}_{\! \perp} - \mbf{P}'_\perp\bigr) . 
\label{eq:Pnorm}
\end{equation}
The momentum generators $P^+$ and
$\mbf{P}_{\! \perp}$ are kinematical; {\em i.e.}, they are independent of
the interactions. The LF time evolution operator  
$P^- = i \frac{d}{d\tau}$ 
can be derived directly from the QCD Lagrangian in the light-cone
gauge $A^+=0$.
In principle, the complete set of bound states and scattering
eigensolutions of $H_{LF}$ can be obtained by solving the light-front
Heisenberg equation $H_{LF} \ket{\psi_h} = {\cal M}^2_h \ket{\psi_h}$,
where $\ket{\psi_h}$ is an expansion in multi-particle Fock eigenstates
$\{\ket{n} \}$ of the free LF Hamiltonian: 
$\vert \psi_h \rangle = \sum_n \psi_{n/h} \vert \psi_h \rangle $.
The LF Heisenberg
equation has in fact been solved for QCD$(1\! + \!1)$ and a number of
other theories using the discretized light-cone quantization method~\cite{Pauli:1985ps}.
The light-cone gauge
has the advantage that all gluon degrees of freedom have physical
polarization and positive metric.  In addition,  orbital angular
momentum has a simple physical interpretation in this
representation. The light-front wavefunctions (LFWFs) $\psi_{n/h}$ provide a
{\it frame-independent } representation of a hadron which relates its quark
and gluon degrees of freedom to their asymptotic hadronic state.

Each hadronic eigenstate $\vert \psi_h \rangle$  is expanded in
a Fock-state complete basis of non-interacting $n$-particle states
$\vert n \rangle$ with an infinite number of components
\begin{equation}
\left\vert \psi_h(P^+,\mbf{P}_{\! \perp}, S_z) \right\rangle = 
\sum_{n,\lambda_i}  \! \int  \! \big[d x_i\big] \!
\left[d^2 \mbf{k}_{\perp i}\right] \psi_{n/h}(x_i,\mbf{k}_{\perp i},\lambda_i) 
\frac{1}{\sqrt{x_i}} 
\bigl\vert n: x_i P^+\!, x_i \mbf{P}_{\! \perp} \! + \! \mbf{k}_{\perp i},\lambda_i \bigr\rangle,
\label{eq:LFWFexp}
\end{equation}
where the sum begins with the valence state; e.g., $n \ge 2$ for mesons. The
coefficients of the  Fock expansion
\begin{equation} \label{eq:LFWF}
\psi_{n/h}(x_i, \mbf{k}_{\perp i},\lambda_i) 
= \bigl\langle n:x_i,\mbf{k}_{\perp i},\lambda_i \big\vert \psi_h\bigr\rangle ,
\end{equation}
are independent of the total momentum $P^+$ and $\mbf{P}_{\! \perp}$ of
the hadron and depend only on the relative partonic coordinates,
the longitudinal momentum fraction $x_i = k_i^+/P^+$,
the relative transverse momentum $\mbf{k}_{\perp i}$,
and $\lambda_i$, the
projection of the constituent's spin along the $z$ direction. 
Thus, given the Fock-projection (\ref{eq:LFWF}), the wavefunction
of a hadron is determined in any frame. The
amplitudes $\psi_{n/h}$ represent the probability amplitudes to find
on-mass-shell constituents $i$ with longitudinal momentum $ x_i P^+$, 
transverse momentum $x_i \mbf{P}_{\! \perp} + \mbf{k}_{\perp i}$,
helicity $\lambda_i$ and invariant mass
\begin{equation}
\mathcal{M}_n^2 = \sum_{i=1}^n k_i^\mu k_{i \mu} =
\sum_{i=1}^n \frac{\mathbf{k}^2_{\perp i} + m_i^2}{x_i},
\label{eq:invM}
\end{equation}
in the hadron $h$. Momentum conservation requires
$\sum_{i=1}^n x_i =1$ and $\sum_{i=1}^n \mbf{k}_{\perp i}=0$.
In addition, each light front wavefunction
$\psi_{n/h}(x_i,\vec k_{\perp i},\lambda_i)$ obeys the angular momentum sum 
rule~\cite{Brodsky:2000ii}
$J^z  = \sum_{i=1}^n  S^z_i + \sum_{i=1}^{n-1} L^z_i $,
where $S^z_i = \lambda_i $ and the $n-1$ orbital angular momenta
have the operator form 
$L^z_i =-i \left(\frac{\partial}{\partial k^x_i}k^y_i -
\frac{\partial}{\partial k^y_i}k^x_i \right)$.
It should be emphasized that the assignment of quark and gluon spin and orbital angular momentum of a hadron is  a gauge-dependent concept. The LF framework in light-cone gauge $A^+=0$ provides a 
physical definition since there are no gauge field ghosts and the gluon 
has spin-projection $J^z= \pm 1$; moreover, it is frame-independent.

The LFWFs are normalized according to
\begin{equation}
\sum_n  \int \big[d x_i\big] \left[d^2 \mbf{k}_{\perp i}\right]
\,\left\vert \psi_{n/h}(x_i, \mbf{k}_{\perp i}) \right\vert^2 = 1,
\label{eq:LFWFnorm}
\end{equation}
where the measure of the constituents phase-space momentum
integration  is
\begin{equation}
\int \big[d x_i\big] \equiv
\prod_{i=1}^n \int dx_i \,\delta \Bigl(1 - \sum_{j=1}^n x_j\Bigr) ,
\end{equation}
\begin{equation}
\int \left[d^2 \mbf{k}_{\perp i}\right] \equiv \prod_{i=1}^n \int
\frac{d^2 \mbf{k}_{\perp i}}{2 (2\pi)^3} \, (16 \pi^3) \,
\delta^{(2)} \negthinspace\Bigl(\sum_{j=1}^n\mbf{k}_{\perp j}\Bigr),
\end{equation}
for the normalization given by (\ref{eq:Pnorm}). 
The spin indices have been suppressed.

Given the light-front wavefunctions $\psi_{n/h},$ one can 
compute a large range of hadron
observables. For example, the valence and sea quark and gluon
distributions which are measured in deep inelastic lepton scattering
are defined from the squares of the LFWFs summed over all Fock
states $n$. Form factors, exclusive weak transition
amplitudes~\cite{Brodsky:1998hn} such as $B\to \ell \nu \pi$,  and
the generalized parton distributions~\cite{Brodsky:2000xy} measured
in deeply virtual Compton scattering are (assuming the ``handbag"
approximation) overlaps of the initial and final LFWFs with $n
=n^\prime$ and $n =n^\prime+2$.   In the case of  deeply virtual meson production such as $\gamma^* p \to \pi X$ and $\gamma^* p \to \rho p$,  the meson enters the amplitude directly through its LFWF. 
In  inclusive reactions such as electron-positron annihilation to jets, the hadronic light-front wavefunctions are the amplitudes which control the coalescence of comoving quarks and gluons into hadrons. Thus one can study hadronization at the amplitude level.  Light-front wavefunctions also control higher-twist contributions to inclusive and 
semi-inclusive reactions~\cite{Berger:1979du,Brodsky:2008}.

The gauge-invariant distribution
amplitude $\phi_H(x_i,Q)$ defined from the integral over the
transverse momenta $\mbf{k}^2_{\perp i} \le Q^2$ of the valence
(smallest $n$) Fock state provides a fundamental measure of the
hadron at the amplitude level~\cite{Lepage:1979zb,Efremov:1979qk};
they  are the nonperturbative input to the factorized form of hard
exclusive amplitudes and exclusive heavy hadron decays in
perturbative QCD. The resulting distributions obey the DGLAP and
ERBL evolution equations as a function of the maximal invariant
mass, thus providing a physical factorization
scheme~\cite{Lepage:1980fj}. In each case, the derived quantities
satisfy the appropriate operator product expansions, sum rules, and
evolution equations. However, at large $x$ where the struck quark is
far-off shell, DGLAP evolution is quenched~\cite{Brodsky:1979qm}, so
that the fall-off of the DIS cross sections in $Q^2$ satisfies
inclusive-exclusive duality at fixed $W^2.$

The holographic mapping of hadronic LFWFs to AdS string modes
is most transparent when one uses the impact parameter space representation.
The total position coordinate of a hadron or its transverse center
of momentum $\mbf{R}_\perp$, is defined in terms of the energy
momentum tensor $T^{\mu \nu}$
\begin{equation}
\mbf{R}_\perp = \frac{1}{P^+} \int dx^- 
\negthinspace \int d^2 \mbf{x}_\perp \,T^{++} \,
\mbf{x}_\perp.
\end{equation}
In terms of partonic transverse coordinates
$x_i \mbf{r}_{\perp i} = x_i \mbf{R}_\perp \! + \! \mbf{b}_{\perp i}$,
where  the $\mbf{r}_{\perp i}$ are the physical transverse position
coordinates and  $\mbf{b}_{\perp i}$ frame independent  internal
coordinates, conjugate to the relative coordinates $\mbf{k}_{\perp i}$. 
Thus, $\sum_i \mbf{b}_{\perp i} = 0$ and  
$\mbf{R}_\perp = \sum_i x_i \mbf{r}_{\perp i}$.
The LFWFs $\psi_n(x_j, \mbf{k}_{\perp j})$ can be expanded in terms of the $n-1$ independent
transverse coordinates $\mbf{b}_{\perp j}$,  $j = 1,2,\dots,n-1$
\begin{equation} \label{eq:LFWFb}
\psi_n(x_j, \mathbf{k}_{\perp j}) =  (4 \pi)^{(n-1)/2} 
\exp{\Big(i \sum_{j=1}^{n-1} \mathbf{b}_{\perp j} \cdot \mbf{k}_{\perp j}\Big)} \,
\widetilde{\psi}_n(x_j, \mathbf{b}_{\perp j}).
\end{equation}
The normalization is defined by
\begin{equation}  
\sum_n  \prod_{j=1}^{n-1} \int d x_j d^2 \mbf{b}_{\perp j} 
\left\vert\widetilde \psi_n(x_j, \mbf{b}_{\perp j})\right\vert^2 = 1.
\end{equation}

\subsection{The Form Factor in Light-Front QCD}

One of the important advantages of the light-front formalism is that current
matrix elements can be represented without approximation as overlaps
of light-front wavefunctions. In the case of the elastic space-like
form factors, the matrix element of the $J^+$ current only couples
Fock states with the same number of constituents.  
It is convenient to choose the light-front frame coordinates
\begin{eqnarray} \label{eq:qframe}
P  &=& (P^+, P^-, \mbf{P}_{\! \perp}) = \Bigl( P^+,\frac{ M^2}{ P^+},
\vec 0_{\perp} \Bigr),\\q  &=& (q^+, q^-, \mbf{q}_{\perp}) = \Bigl(
0,  \frac{-q^2}{P^+}, \mbf{q}_{\perp} \Bigr), \nonumber
\end{eqnarray}
where  $q^2 = - Q^2 = -2 P \cdot q = - \mbf{q}^2_\perp$ is the space-like
four momentum squared transferred to the composite system. The
electromagnetic form factor of a meson is defined in terms of the
hadronic  amplitude of the electromagnetic current evaluated at
light-cone time $x^+ = 0$:
$\left\langle P' \left\vert J^+(0) \right\vert P\right\rangle = 2
\left( P + P'\right)^+ F(Q^2)$,
where $P' = P + q$ and $F(0) = 1$. If
the charged parton $n$ is the active constituent struck by the
current, and the quarks $i = 1,2, \dots ,n-1$ are spectators, then
the Drell-Yan West formula~\cite{Drell:1969km,West:1970av,Brodsky:1980zm} in impact space is
\begin{equation} \label{eq:FFb}
F(q^2) =  \sum_n  \prod_{j=1}^{n-1}\int d x_j d^2 \mbf{b}_{\perp j}
\exp\!{\Bigl(i \mbf{q}_\perp \cdot \sum_{j=1}^{n-1} x_j \mbf{b}_{\perp j}\Bigr)}
\left\vert \widetilde \psi_n(x_j, \mbf{b}_{\perp j})\right\vert^2,
\end{equation}
corresponding to a change of transverse momenta $x_j \mbf{q}_\perp$
for each of the $n-1$ spectators.  This is a convenient form for
comparison with AdS results, since the form factor is expressed in
terms of the product of light-front wave functions with identical
variables.

\section{Light-Front /AdS Duality }

We can now establish an explicit connection between the AdS/CFT and
the LF formulae. To make more transparent the holographic connection
between AdS$_5$ and the conformal quantum field theory defined at
its asymptotic $z\to 0$ boundary, it is convenient to use  the AdS
metric (\ref{eq:AdSzLF}) in terms of light front coordinates $x^\pm =
x^0 \pm x^3$.
It is also useful to express (\ref{eq:FFb}) in terms of an effective
single particle transverse distribution $\widetilde \rho$
~\cite{Brodsky:2006uqa}
\begin{equation} \label{eq:FFzeta}
F(q^2) = 2 \pi \int_0^1 dx \frac{(1-x)}{x}  \int \zeta d \zeta\,
J_0\negthinspace\left(\zeta q \sqrt{\frac{1-x}{x}}\right) \tilde
\rho(x,\zeta),
\end{equation}
where we have introduced the variable
\begin{equation}
\zeta = \sqrt{\frac{x}{1-x}} ~\Big\vert \sum_{j=1}^{n-1} x_j
\mathbf{b}_{\perp j}\Big\vert,
\end{equation}
representing the $x$-weighted transverse impact coordinate of the
spectator system. On the other hand the form
factor in AdS space (\ref{eq:FFAdS}) is represented as the overlap in the fifth
dimension coordinate $z$ of the normalizable modes dual to the
incoming and outgoing hadrons, $\Phi_P$ and $\Phi_{P'}$, with the
non-normalizable source mode, $J(Q,z) = z Q K_1(z Q)$.
If we compare (\ref{eq:FFzeta}) in impact space with the expression
for the form factor in AdS space (\ref{eq:FFAdS}) for arbitrary
values of $Q$ using the identity
\begin{equation} \label{eq:intJ}
\int_0^1 dx \, J_0\negthinspace\left(\zeta Q
\sqrt{\frac{1-x}{x}}\right) = \zeta Q K_1(\zeta Q),
\end{equation}
then we can identify the spectator density function appearing in the
light-front formalism with the corresponding AdS density
\begin{equation} \label{eq:hc}
\tilde \rho(x,\zeta)
 =  \frac{R^3}{2 \pi} \frac{x}{1-x}
\frac{\left\vert \Phi(\zeta)\right\vert^2}{\zeta^4}.
\end{equation}
Equation (\ref{eq:hc}) gives a precise relation between  string
modes $\Phi(\zeta)$ in AdS$_5$ and the QCD transverse charge density
$\tilde\rho(x,\zeta)$. The variable $\zeta$ represents a measure of
the transverse separation between point-like constituents, and it is
also the holographic variable $z$ characterizing the string scale in
AdS. Consequently the AdS string mode $\Phi(z)$ can be regarded as
the probability amplitude to find $n$ partons at transverse impact
separation $\zeta = z$. Furthermore, its eigenmodes determine the
hadronic spectrum~\cite{Brodsky:2006uqa}.
In the case of a two-parton constituent bound state, the
correspondence between the string amplitude $\Phi(z)$ and the
light-front wave function $\widetilde\psi(x,\mathbf{b})$ is
expressed in closed form~\cite{Brodsky:2006uqa}
\begin{equation}  \label{eq:Phipsi}
\left\vert\widetilde\psi(x,\zeta)\right\vert^2 = \frac{R^3}{2 \pi}
~x(1-x)~ \frac{\left\vert \Phi(\zeta)\right\vert^2}{\zeta^4},
\end{equation}
where $\zeta^2 \! = \!x(1-x) \mathbf{b}_\perp^2$. Here $\mbf{b}_\perp$ is
the impact separation conjugate to $\mbf{k}_\perp$.

Hadron form factors can thus be predicted from the overlap integral
of string modes propagating in AdS space with the boundary electromagnetic
source which probes the AdS interior, or equivalently by using the 
Drell-Yan-West formula in physical space-time. If both quantities
represent the same physical observable for any value of the
transfer momentum $q^2$, an exact correspondence can be
established between the string modes $\Phi$ in fifth-dimensional
AdS space and the light-front wavefunctions of hadrons $\psi_{n/h}$
in 3+1 spacetime~\cite{Brodsky:2006uqa}.  
One can thus use holography to
map the functional from of the string modes  $\Phi(z)$ in  AdS space
to the light front wavefunctions in physical  space time by
identifying $z$ with the  transverse variable $ \zeta =\sqrt{x/
(1-x)} |\vec \eta_\perp | .$  Here  $\vec \eta_\perp =
\sum^{n-1}_{i=1}  x_i  \mbf{b}_{\perp i}$ is the weighted impact
separation,  summed over the impact separation of the spectator
constituents.  The leading large-$Q^2$  behavior of form factors in
AdS/QCD arises from small $\zeta \sim 1/ Q$,  corresponding to small
transverse separation.
The  form factor of a hadron  at large $Q^2$ thus arises from the small
$z$ kinematic domain in  AdS space. According to the AdS/CFT
duality, this corresponds to small distances $x^\mu x_\mu \sim
{1/Q^2}$ in physical space-time, the domain where the current matrix
elements are controlled by the conformal twist-dimension, $\Delta$,
of the hadron's interpolating operator.  In the case of the front
form, where $x^+=0$,  this corresponds to small transverse
separation $x^\mu x_\mu =  -\mbf{x}^2_\perp.$

As we have shown, the eigenvalues of the effective light-front equation provide a  good description of the meson and baryon spectra for light
quarks, and its eigensolutions provide a remarkably simple but realistic model of their valence wavefunctions.  The
resulting normalized light-front wavefunctions  for the truncated space model are
\begin{equation} \widetilde \psi_{L,k}(x, \zeta) =  B_{L,k} \sqrt{x(1-x)} J_L \left(\zeta \beta_{L,k} \Lambda_{\rm QCD}\right) \theta\big(\zeta \le \Lambda^{-1}_{\rm
QCD}\big), \end{equation} where 
$B_{L,k} = \pi^{-\frac{1}{2}} {\Lambda_{\rm QCD}/ J_{1+L}(\beta_{L,k})}$. 
The results  display confinement at
large inter-quark separation and conformal symmetry at short distances, thus reproducing dimensional counting rules for hard exclusive
processes. We have also derived analogous equations for baryons composed of massless quarks using a LF Dirac matrix representation for the baryon
system. Most important, the eigensolutions of the AdS/CFT equation can be mapped to light-front equations of the hadrons in physical space-time, thus
providing an elegant description of the light hadrons at the amplitude level.  The meson LFWF is
illustrated in Fig.\ref{fig:MesonLFWF}. 
\begin{figure}[ht]
\centering
\includegraphics[angle=0,width=9.6cm]{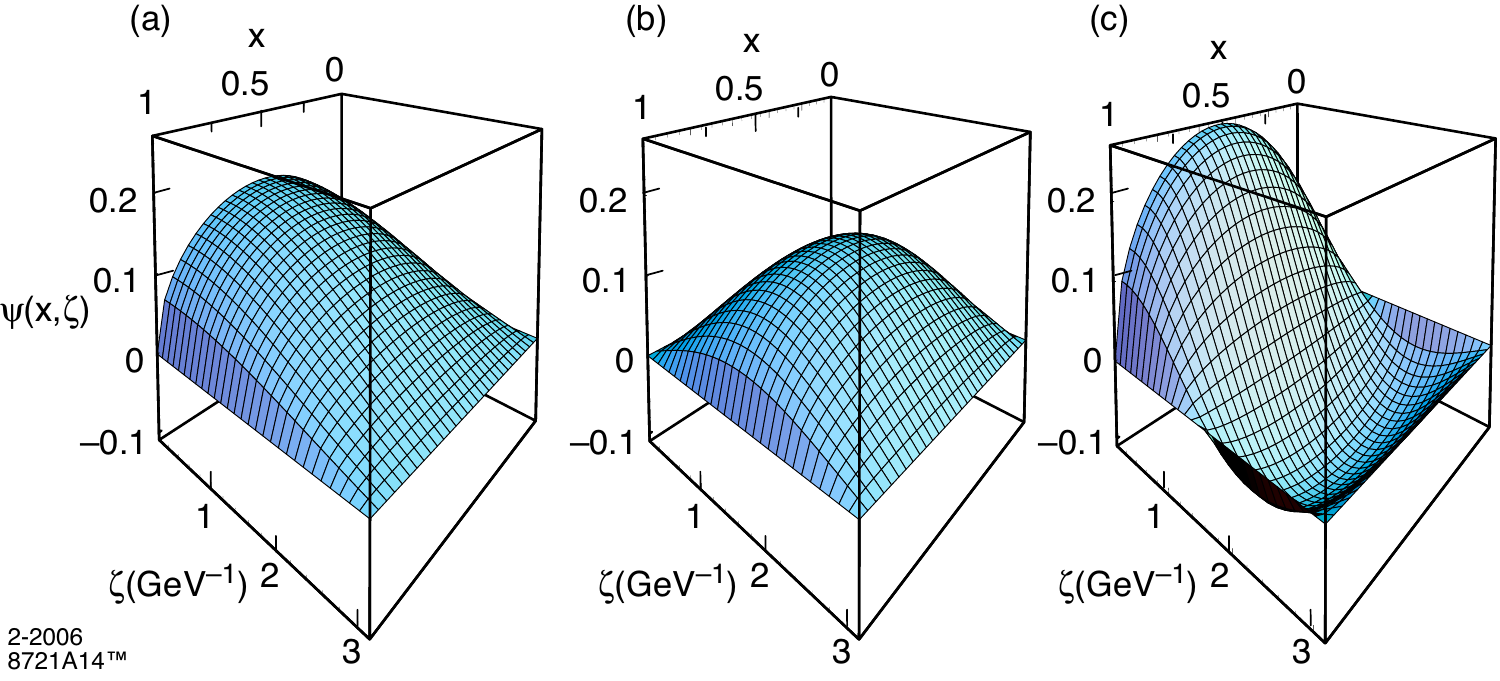}
\caption{AdS/QCD Predictions for the light-front wavefunctions of a
meson in the hard-wall model: (a) $n = 0, L = 0$, (b) $n = 0, L =
1$, (c)  $n = 1, L = 0$.} \label{fig:MesonLFWF}
\end{figure}

\subsection{Light-Front Mapping in the Soft-Wall Model}

As discussed above, in the soft-wall model the current decouples from the dilaton field at large
$Q^2$ and we recover our previous scaling results for the ultraviolet behavior of matrix elements.
To obtain the corresponding basis set of LFWFs we compare the DYW expression of the form
factor  (\ref{eq:FFzeta}) with the AdS form factor (\ref{eq:FFAdSSW})
for large values  of $Q$.
Thus, in the large $Q$ limit we can identify the light-front spectator density
with the corresponding AdS density
\begin{equation} \label{eq:PhirhoSW}
\tilde \rho(x,\zeta)
=   \frac{R^3}{2 \pi} \frac{x}{1-x} \, e^{- \kappa^2 \zeta^2} \,
\frac{ \left\vert \Phi(\zeta)\right\vert^2}{\zeta^4} .
\end{equation}

When summed over all Fock states  the Drell-Yan-West  formula gives an exact result. The
formula describes the coupling of the free electromagnetic current to the elementary constituents
in the interaction representation. In the presence of a dilaton field in AdS space, or in the case
where the
electromagnetic probe propagates in modified confining AdS metrics, the electromagnetic
AdS mode is no longer dual to a the free quark current, but  dual 
to a dressed current, {\it i.e.}, a hadronic electromagnetic current including virtual $\bar q q$
pairs and thus confined. Consequently, at finite values of the momentum transfer $Q^2$ our
simple identification discussed above has to be reinterpreted since we are comparing states in
different representations:  the interaction representation in light-cone QCD versus the Heisenberg representation in AdS. However both quantities should represent the same observables.
We thus expect that the modified mapping corresponds to the presence of higher Fock states in the
hadron.

\section{Holographic Light-Front Wave Functions and Distribution Amplitudes of Flavored
Mesons}

As we have shown above, holographic light-front wave functions (LFWFs) of hadronic bound states  follow from the mapping  to physical space-time of string modes $\Phi(z)$ in AdS$_5$ 
space~\cite{Brodsky:2006uqa}. 
For a two-parton bound state the mapping connects the  transverse impact
variable $\zeta$, the invariant separation between point-like constituents,
identified with the holographic variable $z$,
$ \zeta^2 = z^2 = x(1-x) \mathbf{b}_\perp^2$,
where $\mathbf{b}_\perp$ is the internal transverse position coordinate and
$x$ is the quark momentum fraction. In the soft-wall holographic model, the pion LFWF  in impact space 
in the limit of massless constituents has the simple form~\cite{Brodsky:2007hb}  
\begin{equation} \label{eq:pionLFWFb}
\widetilde\psi(x, \mathbf{b}_\perp)_{\bar q q /\pi} =
\frac{ \kappa}{\sqrt{\pi}} \, \sqrt{x(1-x)} \,
 e^{-\half \kappa^2 x(1-x) \mathbf{b}_\perp^2}.
 \end{equation}
The LFWF (\ref{eq:pionLFWFb}) can also be regarded as
as the solution of a transverse oscillator in the light-front plane~\cite{Brodsky:2007hb}.
The LFWF in $\mathbf{k}_\perp$ space is the Fourier transform
\begin{equation} 
\psi(x, \mathbf{k}_\perp)  =
 \frac{4 \pi}{\kappa \sqrt{x(1-x)}} \,
e^{- \frac{\mathbf{k}_\perp^2}{2 \kappa^2 x(1-x)}}.
\ \label{eq:pionLFWFk}
\end{equation}

A simple generalization of the LFWF (\ref{eq:pionLFWFk}) for massive quarks follows
from the assumption that the momentum space LFWF is a function of the invariant
off-energy shell quantity
\begin{equation}
 \mathcal{M}^2 - \mathcal{E} = \sum_{i=1}^n \frac{\mathbf{k}_{\perp i}^2 + m_i^2}{x_i}.
 \end{equation}
 Thus the holographic soft-wall LFWF ansatz for a meson bound state with massive constituents
 \begin{equation} \label{eq:pionLFWFkm}
 \psi(x, \mathbf{k}_\perp) \sim \frac{4 \pi }{\kappa \sqrt{x(1-x)}} \,
e^{- \frac{1}{2 \kappa^2} \left(\frac{\mathbf{k}_\perp^2}{ x(1-x)}
+ \frac{m_1^2}{x} + \frac{m_2^2}{1-x}\right)}.
\end{equation}
The Fourier transform of (\ref{eq:pionLFWFkm}) is the impact space LFWF
\begin{equation} \label{eq:MLFWFbm}
 \widetilde\psi(x, \mathbf{b}_\perp)  \sim
\frac{ \,\kappa}{\sqrt{\pi}} \, \sqrt{x(1-x)} \,
 e^{-\half \kappa^2  x(1-x) \mathbf{b}_\perp^2  - \frac{1} {\kappa^2}
\left[\frac{m_1^2}{x} + \frac{m_2^2}{1-x}\right]}.
 \end{equation}
Impact space holographic LFWFs for the $\pi$, $K$, $D$, $\eta_c$, $B$ and $\eta_b$ mesons are depicted in
Fig. \ref{fig:MLFWF}.
\begin{figure}[htbp]
\begin{center}
\includegraphics[angle=0,width=6.6cm]{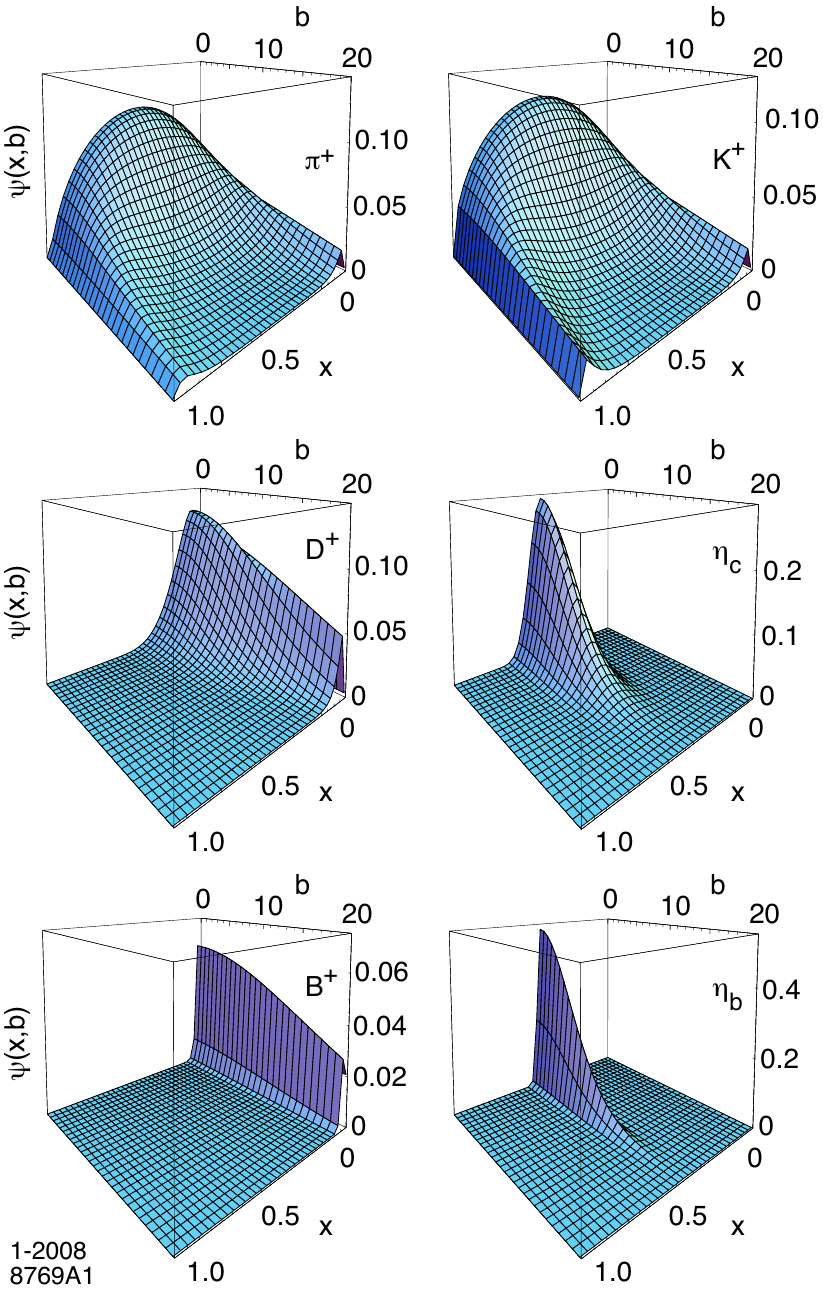}
\caption{Two-parton flavored meson holographic LFWF $\psi(x, \mathbf{b}_\perp)$:
$\vert \pi^+ \rangle = \vert u \bar d \rangle$,
$\vert K^+\rangle = \vert u \bar s \rangle$,  
$\vert D^+ \rangle = \vert c \bar d \rangle$, 
$\vert \eta_c \rangle = \vert c \bar c \rangle$,
$\vert B^+ \rangle = \vert  u \bar b \rangle$  and 
 $\vert  \eta_b \rangle = \vert b \bar b \rangle$.
Values for the quark masses used are $m_u = 2$ MeV, $m_d = 5$ MeV,  $m_s = 95$ Mev, $m_c =  1.25$ GeV and $m_b = 4.2$ GeV. The value of $\kappa= 0. 375$ GeV is  from the pion form 
factor~\cite{Brodsky:2007hb}. }
\label{fig:MLFWF}
\end{center}
\end{figure}

The  non-perturbative input to hard
exclusive processes and heavy hadron decays can be computed in terms of gauge invariant hadronic distribution
amplitudes (DAs), which describe the momentum-fraction distribution of partons at zero transverse
impact distance in a Fock state with a fixed number of constituents, and thus they involve
current or Lagrangian quark masses in the light-front wave function. The meson DA
is computed from the transverse integral of the valence quark light-front wavefunction in
the light-cone gauge~\cite{Lepage:1980fj} 
\begin{equation}
\phi_M(x,Q) = \int^{\mathbf{k}_\perp^2 < Q^2} \! \frac{d^2 \mathbf{k}_\perp}{16 \pi^3} \, \psi_M(x, \mathbf{k}_\perp),
\end{equation}
and thus  $\phi(x) \equiv \phi(x, Q \to \infty) \to 
\widetilde \psi(x,\mathbf{b}_\perp \to 0)/\sqrt{4 \pi}$.  
From (\ref{eq:MLFWFbm}) we obtain the holographic distribution amplitude $\phi(x)$
\begin{equation} \label{eq:DA}
 \phi_M(x)  \sim
\frac{ \,\kappa}{2 \pi} \, \sqrt{x(1-x)} \, 
 e^{- \frac{1} {2 \kappa^2}
\left[\frac{m_1^2}{x} + \frac{m_2^2}{1-x}\right]} ,
 \end{equation}
in the soft wall model. Predictions for the first and second moment of the meson distribution amplitude
\begin{equation}
\langle \xi^N \rangle_M = \frac{\int_{-1}^{1} \xi^N \phi_M(\xi)}{ \int_{-1}^{1} \phi_M(\xi)},
\end{equation}
and comparisons with available lattice computations 
are given on Table {\ref{tab:MDA} . 
\begin{table}[htdp]
\centering
\begin{minipage}{\textwidth}
\centering
\caption{Predictions for first and second moment of meson DA (top) and comparisons with available lattice results (bottom).
 Values of quark masses and $\kappa$ as 
in Fig. \ref{fig:MLFWF}.}
\begin{tabular}{ccc}
\hline\hline \vspace{2pt}
$M$ & $\langle \xi \rangle_M$ &  $\langle \xi^2 \rangle_M$\\
\hline
$\pi$ &  & 0.25 \\
$K$ & $0.04 \pm 0.02\,^a$  
& $0.235 \pm 0.005$\footnote{The results correspond to  $m_s = 65 \pm 25$  MeV from~\cite{Yao:2006px}.}\\
$D$ & 0.71 & 0.54 \\
$\eta_c$ &  & 0.02 \\
$B$ &  0.96 & 0.91\\
$\eta_b$ &  & 0.002\\
\hline \hline
$\pi$ &    &  $0.28 \pm 0.03 \footnote{Lattice results from Ref.~\cite{Donnellan:2007xr}} $ \\
$K$ &  $0.029 \pm 0.002\,^b $   &  $0.27 \pm0.02\,^b$\\
$\pi$ &    &   $0.269 \pm 0.039 \footnote{Lattice results from Ref.~\cite{Braun:2006dg}}$ \\
$K$ & $ 0.0272 \pm0.0005\,^c$    & $0.260 \pm 0.006\,^c$
\end{tabular}
\end{minipage}
\label{tab:MDA}
\end{table}

It is interesting to note that the pion distribution amplitude
predicted by AdS/QCD 
has a quite different $x$-behavior than the
asymptotic distribution amplitude predicted from the PQCD
evolution~\cite{Lepage:1979zb} of the pion distribution amplitude.
In the chiral limit, the AdS distribution amplitude $\phi_{\rm AdS}(x) \sim \sqrt{x(1-x)}$  gives for the second moment $\langle \xi^2 \rangle_{\rm AdS} \to 1/4$, compared with the asymptotic  value $\langle \xi^2 \rangle_{\rm PQCD} \to 1/5$ from the PQCD asymptotic DA  $\phi_{\rm PQCD}(x) \sim x(1-x)$.
This observation appears to be consistent with the results of the Fermilab diffractive dijet experiment~\cite{Aitala:2000hb} which shows a broader  $x$ distribution for the dijets at small transverse momentum $k_\perp \le 1 $ GeV.  The broader
shape of the pion distribution increases the magnitude of the
leading twist perturbative QCD prediction for the pion form factor
by a factor of $16/9$ compared to the prediction based on the
asymptotic form, bringing the PQCD prediction  close to the
empirical pion form factor~\cite{Choi:2006ha}.

Since they are complete and orthonormal, the AdS/CFT model
wavefunctions can be used as an initial ansatz for a variational
treatment or  the basis states for the diagonalization of the light-front
QCD Hamiltonian $H^{QCD}_{LF}$~\cite{Brodsky:1997de}.   
Even if one restricts the proton basis to $\vert uud \rangle, 
\vert uudgg\rangle$, and $\vert uudq\bar q\rangle $ Fock states, the resulting eigensolution will contain the effects of gluon exchange, the lowest order contribution to the QCD running coupling, intrinsic gluons and sea quarks.

\section{Conclusions}

One of the key difficulties in studies of quantum chromodynamics has been the absence of an analytic first approximation to the theory which not only can  reproduce the hadronic spectrum, but also provides a good description of hadron wavefunctions.  The AdS/CFT  correspondence provides an elegant semi-classical approximation to QCD, which incorporates both color confinement and the conformal short-distance behavior appropriate for a theory with an infrared fixed point.   Since the hadronic solutions are controlled by their twist dimension $z^\tau$ at small $z$, one  also reproduces dimensional counting rules for hard exclusive processes. 
The AdS/CFT approach leads to a model of hadrons
which has both confinement at large distances and the conformal
scaling properties which reproduce dimensional counting rules for
hard exclusive reactions.  The fundamental equations of AdS/CFT for mesons have
the appearance of a radial Schr\"odinger Coulomb equation, but they are
relativistic, covariant, and analytically tractable. The eigenvalues of the AdS/CFT equations provide a  good description of the meson and baryon
spectra for light quarks~\cite{deTeramond:2005su, deTeramond:2006xb, Brodsky:2006uq, Brodsky:2005kc}, and its eigensolutions provide a remarkably simple but realistic model of their valence
wavefunctions.   One can also derive analogous equations for baryons composed of massless quarks using a Dirac matrix representation for the baryon system~\cite{Brodsky:2007pt}. 

The lowest stable
state of the AdS equations are determined by the Breitenlohner-Freedman bound~\cite{Breitenlohner:1982jf}.  We can model confinement  by imposing Dirichlet boundary
conditions at $\phi(z = 1/\Lambda_{\rm QCD}) = 0.$ The eigenvalues are then given in terms of the roots of the Bessel functions:
$\mathcal{M}_{L,k} = \beta_{L,k} \Lambda_{\rm QCD}$.   Alternatively, one can  introduce a dilaton 
field~\cite{Karch:2006pv} which provides a confinement potential $-\kappa^2 \zeta^2$ to the effective
potential $V(\zeta).$  The resulting hadron spectra are given by linear Regge trajectories in the square
of the hadron masses $\mathcal{M}^2$, characteristic of the Nambu string model.  
The AdS/CFT equations are integrable, and thus the radial and orbital excitations can be obtained from ladder operators~\cite{Brodsky:2007pt}. We have  found that the equations describing 
the propagation of light-front
eigenmodes in 3+1 space possess remarkable algebraic structures. We have also shown
that a simple extension of the conformal algebraic structure is equivalent to the soft-wall model.
For fermionic modes it corresponds to a linear confining potential in the holographic variable $z$.

In this work we have shown that the eigensolutions  $\Phi_H(z)$  of the AdS/CFT equations in the fifth dimension $z$ have a remarkable mapping to the light-front wavefunctions 
$\psi_H(x_i,\mathbf{b}_{\perp i})$, the hadronic amplitudes which describe the valence constituents of hadrons in physical space time, but at fixed light-cone time $\tau=t+z/c=0$.   Similarly,  the AdS/CFT equations for hadrons can be mapped to equivalent light-front equations. 
The correspondence of AdS/CFT amplitudes to the QCD wavefunctions in light-front coordinates in 
physical space-time provides an exact holographic mapping at all energy scales between string modes in AdS space and hadronic boundary states.
Most important, the eigensolutions of the AdS/CFT equation can be mapped to light-front equations of the hadrons in physical space-time, thus
providing an elegant description of the light hadrons at the amplitude level. 

The mapping of AdS/CFT string modes to light-front wave functions thus provides a remarkable analytic  first approximation to QCD.
Since they are complete and orthonormal, the AdS/CFT model
wavefunctions can also be used as a basis for the diagonalization of
the full light-front QCD Hamiltonian, thus systematically improving
the AdS/CFT approximation. 

We have also shown the correspondence between the expressions for current matrix elements in AdS/CFT with the corresponding expressions for form factors as given in the light-front formalism. In first approximation, where one takes the current propagating in a non-confining background, one obtains the Drell-Yan West formula for valence Fock states, corresponding to the interaction picture of the light-front theory.  Hadron form factors can thus be directly predicted from the overlap integrals in AdS space or equivalently by using
the Drell-Yan-West formula in physical space-time.   The form factor at high $Q^2$ receives its main contributions from small $\zeta \sim {1/ Q}$,
corresponding to small $\vec b_\perp = {\cal O}(1/Q)$ and  $1-x  = {\cal O}(1/Q)$.
We have also shown how to improve this approximation  by studying the propagation of non-normalizable solutions representing the electromagnetic  current in a modified AdS 
confining metric, or equivalently in a dilaton background.   This improvement in the description of the current corresponds  in the light-front to multiple hadronic Fock states.  The introduction of the confined current implies that the timelike form factors of  hadrons will be mediated by vector mesons, including radial excitations.  The wavefunction of the normalizable vector meson states $\mathcal{A}(z)$ appearing in the spectral
decomposition of the Green's function, which is dual to the non-normalizable photon propagation mode in AdS, is twist-3~\cite{Brodsky:2007hb}. This is the expected result for even parity axial mesons in QCD, or $L=1$ odd parity vector mesons composed of a scalar squark and anti-squarks.  In the case of quark-antiquark states, one also expects to find  $C= - 1$, twist-2 meson solutions for the zero helicity component of the $\rho$ with $S=1$ and $L=0$, which is supposed to give a dominant contribution to the $\rho$ form factor. 

We have applied our formulation to both the spacelike and timelike pion form factor.  The description of the pion form factor in the spacelike domain is in  good agreement with experiment in both confinement models, the hard and the soft wall holographic models.  In the soft wall model the time-like pion form factor exhibits a pole at the $\rho$ mass with zero width since hadrons are stable in this theory.  If one
introduces  a width, the height of the $\rho$ pole is in reasonable agreement with experiment. 
The space-like Dirac form factor for the proton is also very well reproduced by the double-pole analytic 
expression given in Appendix D of Ref.~\cite{Brodsky:2007hb}  for the case $N=3$. 

The deeply virtual Compton amplitude in the  handbag approximation can be expressed as overlap of light-front wavefunctions~\cite{Brodsky:2000xy}.
The deeply virtual Compton amplitudes can be Fourier transformed to $b_\perp$ and $\sigma = x^-P^+/2$ space providing new
insights into QCD distributions~\cite{Burkardt:2005td, Ji:2003ak, Ralston:2001xs, Brodsky:2006in, Hoyer:2006xg}. The distributions in the light-front direction $\sigma$
typically display diffraction patterns arising from the interference of the initial and final state LFWFs
~\cite{Brodsky:2006in,Brodsky:2006ku}. 
All of these processes  can provide detailed tests of  the AdS/CFT LFWFs predictions.

The phenomenology of the AdS/CFT model is just beginning, but it can
be anticipated that it will have extensive applications to QCD phenomena.
For example, the model LFWFs  obtained from AdS/QCD provide a basis for understanding
hadron structure functions and fragmentation functions as well as higher-twist contributions to inclusive processes at the
amplitude level; the same wavefunctions can describe hadron
formation from the coalescence of co-moving quarks.  The spin and orbital angular momentum
correlations which underly single and double spin correlations are
also described by the AdS/CFT eigensolutions.  The AdS/CFT hadronic
wavefunctions also provide predictions for the generalized parton distributions of hadrons
and their weak decay amplitudes from first principles.  The
amplitudes relevant to diffractive reactions could also be computed.  We
also anticipate that the extension of the AdS/CFT formalism to heavy
quarks will allow a great variety of heavy hadron phenomena to be
analyzed from first principles.

\vspace{15pt}

{\bf \large Acknowledgments}

\vspace{5pt}

SJB thanks Professors A. Zichichi and G. 't Hooft for the opportunity to lecture at this school,  and he is grateful for the intense interest of the students of the school in AdS/QCD developments.
This research was supported by the Department
of Energy contract DE--AC02--76SF00515. The research of GdT is supported in part
by an Aportes grant from Florida Ice \& Farm.


\begin{thebibliography}{0}


 
  \bibitem{vonSmekal:1997is}
  L.~von Smekal, R.~Alkofer and A.~Hauck,
  ``The infrared behavior of gluon and ghost propagators in Landau gauge QCD,''
  Phys.\ Rev.\ Lett.\  {\bf 79}, 3591 (1997)
  [arXiv:hep-ph/9705242].
  %%CITATION = PRLTA,79,3591;%%

\bibitem{Zwanziger:2003cf}
  D.~Zwanziger,
  ``Non-perturbative Faddeev-Popov formula and infrared limit of QCD,''
  Phys.\ Rev.\  D {\bf 69}, 016002 (2004)
  [arXiv:hep-ph/0303028].
  %%CITATION = PHRVA,D69,016002;%%
  
  \bibitem{Alkofer:2004it}
  R.~Alkofer, C.~S.~Fischer and F.~J.~Llanes-Estrada,
  ``Vertex functions and infrared fixed point in Landau gauge SU(N)  Yang-Mills
  theory,''
  Phys.\ Lett.\  B {\bf 611}, 279 (2005)
  [arXiv:hep-th/0412330].
  %%CITATION = PHLTA,B611,279;%%
  
  \bibitem{Fischer:2006vf}
  C.~S.~Fischer and J.~M.~Pawlowski,
  ``Uniqueness of infrared asymptotics in Landau gauge Yang-Mills theory,''
  Phys.\ Rev.\  D {\bf 75}, 025012 (2007)
  [arXiv:hep-th/0609009].
  %%CITATION = PHRVA,D75,025012;%%
  
  \bibitem{Epple:2006hv}
  D.~Epple, H.~Reinhardt and W.~Schleifenbaum,
  ``Confining Solution of the Dyson-Schwinger Equations in Coulomb Gauge,''
  Phys.\ Rev.\  D {\bf 75}, 045011 (2007)
  [arXiv:hep-th/0612241].
  %%CITATION = PHRVA,D75,045011;%%
  
   \bibitem{Kellermann:2008iw}
  C.~Kellermann and C.~S.~Fischer,
  ``The running coupling from the four-gluon vertex in Landau gauge Yang-Mills
  theory,''
  arXiv:0801.2697 [hep-ph].
  %%CITATION = ARXIV:0801.2697;%%

 \bibitem{Alkofer:2008jy}
  R.~Alkofer, M.~Q.~Huber and K.~Schwenzer,
  ``On infrared singularities in Landau gauge Yang-Mills theory,''
  arXiv:0801.2762 [hep-th].
  %%CITATION = ARXIV:0801.2762;%%
  
 \bibitem{Mattingly:1993ej}
  A.~C.~Mattingly and P.~M.~Stevenson,
  ``Optimization Of R(E+ E-) And 'Freezing' Of The QCD Couplant At
  Low-Energies,''
  Phys.\ Rev.\  D {\bf 49}, 437 (1994)
  [arXiv:hep-ph/9307266].
  %%CITATION = PHRVA,D49,437;%%
  

\bibitem{Brodsky:2002nb}
  S.~J.~Brodsky, S.~Menke, C.~Merino and J.~Rathsman,
  ``On the behavior of the effective QCD coupling alpha(tau)(s) at low
  scales,''
  Phys.\ Rev.\  D {\bf 67}, 055008 (2003)
  [arXiv:hep-ph/0212078].
  %%CITATION = PHRVA,D67,055008;%%

\bibitem{Baldicchi:2002qm}
  M.~Baldicchi and G.~M.~Prosperi,
  ``Infrared behavior of the running coupling constant and bound states in
  QCD,''
  Phys.\ Rev.\  D {\bf 66}, 074008 (2002)
  [arXiv:hep-ph/0202172].
  %%CITATION = PHRVA,D66,074008;%%
  
\bibitem{Brodsky:1998ua}
  S.~J.~Brodsky, J.~R.~Pelaez and N.~Toumbas,
  ``Testing {QCD} with hypothetical tau leptons,''
  Phys.\ Rev.\  D {\bf 60}, 037501 (1999)
  [arXiv:hep-ph/9810424].
  %%CITATION = PHRVA,D60,037501;%%
  
   \bibitem{Deur:2005cf}
  A.~Deur, V.~Burkert, J.~P.~Chen and W.~Korsch,
  ``Experimental determination of the effective strong coupling constant,''
  Phys.\ Lett.\  B {\bf 650}, 244 (2007)
  [arXiv:hep-ph/0509113].
  %%CITATION = PHLTA,B650,244;%%

\bibitem{Furui:2006py}
  S.~Furui and H.~Nakajima,
  ``Infrared features of unquenched finite temperature lattice Landau gauge
  QCD,''
  Phys.\ Rev.\  D {\bf 76}, 054509 (2007)
  [arXiv:hep-lat/0612009].
  %%CITATION = PHRVA,D76,054509;%%
  
  \bibitem{Appelquist:2007hu}
  T.~Appelquist, G.~T.~Fleming and E.~T.~Neil,
  ``Lattice Study of the Conformal Window in QCD-like Theories,''
  arXiv:0712.0609 [hep-ph].
  %%CITATION = ARXIV:0712.0609;%%

\bibitem{Antonov:2007mx}
  D.~Antonov and H.~J.~Pirner,
  ``Following gluonic world lines to find the QCD coupling in the infrared,''
  Eur.\ Phys.\ J.\  C {\bf 51}, 633 (2007)
  [arXiv:hep-ph/0702227].
  %%CITATION = EPHJA,C51,633;%%

\bibitem{Brodsky:2007pt}
  S.~J.~Brodsky and G.~F.~de Teramond,
  ``AdS/CFT and QCD,''
  arXiv:hep-th/0702205.
  %%CITATION = HEP-TH/0702205;%%
  
  \bibitem{Cornwall:1981zr}
  J.~M.~Cornwall,
  ``Dynamical Mass Generation In Continuum QCD,''
  Phys.\ Rev.\  D {\bf 26}, 1453 (1982).
  %%CITATION = PHRVA,D26,1453;%%

\bibitem{Parisi:1972zy}
  G.~Parisi,
  ``Conformal invariance in perturbation theory,''
  Phys.\ Lett.\  B {\bf 39}, 643 (1972).
  %%CITATION = PHLTA,B39,643;%%
  

\bibitem{Brodsky:1985ve}
  S.~J.~Brodsky, Y.~Frishman and G.~P.~Lepage,
  ``On The Application Of Conformal Symmetry To Quantum Field Theory,''
  Phys.\ Lett.\  B {\bf 167}, 347 (1986).
  %%CITATION = PHLTA,B167,347;%%

\bibitem{Brodsky:1974vy}
  S.~J.~Brodsky and G.~R.~Farrar,
  ``Scaling Laws For Large Momentum Transfer Processes,''
  Phys.\ Rev.\  D {\bf 11}, 1309 (1975).
  %%CITATION = PHRVA,D11,1309;%%

\bibitem{Matveev:1973ra}
  V.~A.~Matveev, R.~M.~Muradian and A.~N.~Tavkhelidze,
  ``Automodellism in the large - angle elastic scattering and structure of
  hadrons,''
  Lett.\ Nuovo Cim.\  {\bf 7}, 719 (1973).
  %%CITATION = NCLTA,7,719;%%

\bibitem{Brodsky:1989pv}
  S.~J.~Brodsky and G.~P.~Lepage,
  ``Exclusive Processes In Quantum Chromodynamics,''
  Adv.\ Ser.\ Direct.\ High Energy Phys.\  {\bf 5}, 93 (1989).
  %%CITATION = 00319,5,93;%%

\bibitem{Lepage:1980fj}
  G.~P.~Lepage and S.~J.~Brodsky,
  ``Exclusive Processes In Perturbative Quantum Chromodynamics,''
  Phys.\ Rev.\  D {\bf 22}, 2157 (1980).
  %%CITATION = PHRVA,D22,2157;%%

\bibitem{Diehl:2004cx}
  M.~Diehl, T.~Feldmann, R.~Jakob and P.~Kroll,
  ``Generalized parton distributions from nucleon form factor data,''
  Eur.\ Phys.\ J.\  C {\bf 39}, 1 (2005)
  [arXiv:hep-ph/0408173].
  %%CITATION = EPHJA,C39,1;%%

\bibitem{Holt:1990ze}
  R.~J.~Holt,
  ``Exclusive photonuclear reactions and asymptotic scaling,''
  Phys.\ Rev.\  C {\bf 41}, 2400 (1990).
  %%CITATION = PHRVA,C41,2400;%%

\bibitem{Bochna:1998ca}
  C.~Bochna {\it et al.}  [E89-012 Collaboration],
  ``Measurements of deuteron photodisintegration up to 4.0-GeV,''
  Phys.\ Rev.\ Lett.\  {\bf 81}, 4576 (1998)
  [arXiv:nucl-ex/9808001].
  %%CITATION = PRLTA,81,4576;%%

\bibitem{Rossi:2004qm}
  P.~Rossi {\it et al.}  [CLAS Collaboration],
  ``Onset of asymptotic scaling in deuteron photodisintegration,''
  arXiv:hep-ph/0405207.
  %%CITATION = PHRVA,LETT.,012301;%%

\bibitem{Danagoulian:2007gs}
  A.~Danagoulian {\it et al.}  [Hall A Collaboration],
  ``Compton scattering cross section on the proton at high momentum transfer,''
  Phys.\ Rev.\ Lett.\  {\bf 98}, 152001 (2007)
  [arXiv:nucl-ex/0701068].
  %%CITATION = PRLTA,98,152001;%%

\bibitem{Chen:2001sm}
  A.~Chen,
  ``Proton Anti-Proton Pair Production In Two-Photon Collisions At Belle,''
%\href{http://www.slac.stanford.edu/spires/find/hep/www?irn=5539323}{SPIRES entry} 
International Conference on the Structure and Interactions of the Photon and 14th International Workshop on Photon-Photon Collisions (Photon 2001), Ascona, Switzerland, 2-7 Sep 2001.

\bibitem{Chen:2006ug}
  A.~E.~Chen  [BELLE Collaboration],
  ``Two Photon Physics At Belle,''
  Int.\ J.\ Mod.\ Phys.\  A {\bf 21}, 5543 (2006).
  %%CITATION = IMPAE,A21,5543;%%

\bibitem{Maldacena:1997re}
  J.~M.~Maldacena,
  ``The large N limit of superconformal field theories and supergravity,''
  Adv.\ Theor.\ Math.\ Phys.\  {\bf 2}, 231 (1998)
  [Int.\ J.\ Theor.\ Phys.\  {\bf 38}, 1113 (1999)]
  [arXiv:hep-th/9711200].
  %%CITATION = IJTPB,38,1113;%%
  
    \bibitem{Gubser:1998bc}
  S.~S.~Gubser, I.~R.~Klebanov and A.~M.~Polyakov,
  ``Gauge theory correlators from non-critical string theory,''
  Phys.\ Lett.\ B {\bf 428}, 105 (1998)
  [arXiv:hep-th/9802109].
  %%CITATION = HEP-TH 9802109;%%
  
\bibitem{Witten:1998qj}
  E.~Witten,
  ``Anti-de Sitter space and holography,''
  Adv.\ Theor.\ Math.\ Phys.\  {\bf 2}, 253 (1998)
  [arXiv:hep-th/9802150].
  %%CITATION = HEP-TH 9802150;%%

\bibitem{Polchinski:2001tt}
  J.~Polchinski and M.~J.~Strassler,
  ``Hard scattering and gauge/string duality,''
  Phys.\ Rev.\ Lett.\  {\bf 88}, 031601 (2002)
  [arXiv:hep-th/0109174].
  %%CITATION = PRLTA,88,031601;%%
  
  \bibitem{Karch:2006pv}
  A.~Karch, E.~Katz, D.~T.~Son and M.~A.~Stephanov,
  ``Linear confinement and AdS/QCD,''
  Phys.\ Rev.\  D {\bf 74}, 015005 (2006)
  [arXiv:hep-ph/0602229].
  %%CITATION = PHRVA,D74,015005;%%

\bibitem{Brodsky:2003px}
  S.~J.~Brodsky and G.~F.~de Teramond,
  ``Light-front hadron dynamics and AdS/CFT correspondence,''
  Phys.\ Lett.\  B {\bf 582}, 211 (2004)
  [arXiv:hep-th/0310227].
  %%CITATION = PHLTA,B582,211;%%

\bibitem{Janik:1999zk}
  R.~A.~Janik and R.~Peschanski,
  ``High energy scattering and the AdS/CFT correspondence,''
  Nucl.\ Phys.\  B {\bf 565}, 193 (2000)
  [arXiv:hep-th/9907177].
  %%CITATION = NUPHA,B565,193;%%
  
  \bibitem{Lin:2006rf}
  S.~Lin and E.~Shuryak,
  ``Toward the AdS/CFT gravity dual for High Energy Collisions: I. Falling into
  the AdS,''
  arXiv:hep-ph/0610168;
  %%CITATION = HEP-PH/0610168;%%
  % \bibitem{Lin:2007fa}
  %S.~Lin and E.~Shuryak,
  ``Toward the AdS/CFT Gravity Dual for High Energy Collisions: II. The Stress
  Tensor on the Boundary,''
  arXiv:0711.0736 [hep-th].
  %%CITATION = ARXIV:0711.0736;%%
  
\bibitem{Brower:2007qh}
  R.~C.~Brower, M.~J.~Strassler and C.~I.~Tan,
  ``On the Eikonal Approximation in AdS Space,''
  arXiv:0707.2408 [hep-th].
  %%CITATION = ARXIV:0707.2408;%%
  
\bibitem{Penedones:2007ns}
  J.~Penedones,
  ``High Energy Scattering in the AdS/CFT Correspondence,''
  arXiv:0712.0802 [hep-th].
  %%CITATION = ARXIV:0712.0802;%%
  
  \bibitem{Erlich:2005qh}
  J.~Erlich, E.~Katz, D.~T.~Son and M.~A.~Stephanov,
  ``QCD and a holographic model of hadrons,''
  Phys.\ Rev.\ Lett.\  {\bf 95}, 261602 (2005)
  [arXiv:hep-ph/0501128].
  %%CITATION = PRLTA,95,261602;%%

 \bibitem{Boschi-Filho:2002vd}
  H.~Boschi-Filho and N.~R.~F.~Braga,
  ``Gauge / string duality and scalar glueball mass ratios,''
  JHEP {\bf 0305}, 009 (2003)
  [arXiv:hep-th/0212207].
  %%CITATION = HEP-TH 0212207;%%

\bibitem{deTeramond:2004qd}
  G.~F.~de Teramond and S.~J.~Brodsky,
  ``Baryonic states in QCD from gauge / string duality at large N(c),''
  arXiv:hep-th/0409074.
  %%CITATION = HEP-TH 0409074;%%

\bibitem{deTeramond:2005su}
  G.~F.~de Teramond and S.~J.~Brodsky,
  ``The hadronic spectrum of a holographic dual of QCD,''
  Phys.\ Rev.\ Lett.\  {\bf 94}, 201601 (2005)
  [arXiv:hep-th/0501022].
  %%CITATION = HEP-TH 0501022;%%
  
   \bibitem{Brodsky:2006uqa}
  S.~J.~Brodsky and G.~F.~de Teramond,
  ``Hadronic spectra and light-front wavefunctions in holographic QCD,''
  Phys.\ Rev.\ Lett.\  {\bf 96}, 201601 (2006)
  [arXiv:hep-ph/0602252].
  %%CITATION = PRLTA,96,201601;%%
    
  \bibitem{Evans:2006ea}
  N.~Evans and A.~Tedder,
  ``Perfecting the ultra-violet of holographic descriptions of QCD,''
  Phys.\ Lett.\  B {\bf 642}, 546 (2006)
  [arXiv:hep-ph/0609112].
  %%CITATION = PHLTA,B642,546;%%

  \bibitem{Hong:2006ta}
  D.~K.~Hong, T.~Inami and H.~U.~Yee,
  ``Baryons in AdS/QCD,''
  Phys.\ Lett.\  B {\bf 646}, 165 (2007)
  [arXiv:hep-ph/0609270].
  %%CITATION = PHLTA,B646,165;%%
  
  \bibitem{Colangelo:2007pt}
  P.~Colangelo, F.~De Fazio, F.~Jugeau and S.~Nicotri,
  ``On the light glueball spectrum in a holographic description of QCD,''
  Phys.\ Lett.\  B {\bf 652}, 73 (2007)
  [arXiv:hep-ph/0703316].
  %%CITATION = PHLTA,B652,73;%%
  
  \bibitem{Forkel:2007ru}
  H.~Forkel,
  ``Holographic glueball structure,''
  arXiv:0711.1179 [hep-ph].
  %%CITATION = ARXIV:0711.1179;%%
  
  \bibitem{DaRold:2005zs}
  L.~Da Rold and A.~Pomarol,
   ``Chiral symmetry breaking from five dimensional spaces,''
  Nucl.\ Phys.\ B {\bf 721}, 79 (2005)
  [arXiv:hep-ph/0501218];
  %%CITATION = HEP-PH 0501218;%%
  ``The scalar and pseudoscalar sector in a five-dimensional approach to chiral
  symmetry breaking,''
  JHEP {\bf 0601}, 157 (2006)
  [arXiv:hep-ph/0510268].
  %%CITATION = HEP-PH 0510268;%%

  \bibitem{Hirn:2005nr}
  J.~Hirn and V.~Sanz,
   ``Interpolating between low and high energy QCD via a 5D Yang-Mills  model,''
  JHEP {\bf 0512}, 030 (2005)
  [arXiv:hep-ph/0507049];
  %%CITATION = HEP-PH 0507049;%%
 %\bibitem{Hirn:2005vk}
  J.~Hirn, N.~Rius and V.~Sanz,
  ``Geometric approach to condensates in holographic QCD,''
  Phys.\ Rev.\ D {\bf 73}, 085005 (2006)
  [arXiv:hep-ph/0512240].
  %%CITATION = HEP-PH 0512240;%%

  \bibitem{Ghoroku:2005vt}
  K.~Ghoroku, N.~Maru, M.~Tachibana and M.~Yahiro,
  ``Holographic model for hadrons in deformed AdS(5) background,''
  Phys.\ Lett.\ B {\bf 633}, 602 (2006)
  [arXiv:hep-ph/0510334].
  %%CITATION = HEP-PH 0510334;%%
  
  \bibitem{Krikun:2008tf}
  A.~Krikun,
  ``On two-point correlation functions in AdS/QCD,''
  arXiv:0801.4215 [hep-th].
  %%CITATION = ARXIV:0801.4215;%%
  
  \bibitem{BoschiFilho:2005mw}
  H.~Boschi-Filho, N.~R.~F.~Braga and C.~N.~Ferreira,
  ``Static strings in Randall-Sundrum scenarios and the quark anti-quark
  potential,''
  Phys.\ Rev.\  D {\bf 73}, 106006 (2006)
  [Erratum-ibid.\  D {\bf 74}, 089903 (2006)]
  [arXiv:hep-th/0512295].
  %%CITATION = PHRVA,D73,106006;%%
  
  \bibitem{Andreev:2006ct}
  O.~Andreev and V.~I.~Zakharov,
  ``Heavy-quark potentials and AdS/QCD,''
  Phys.\ Rev.\  D {\bf 74}, 025023 (2006)
  [arXiv:hep-ph/0604204].
  %%CITATION = PHRVA,D74,025023;%%
  
  \bibitem{Hambye:2005up}
  T.~Hambye, B.~Hassanain, J.~March-Russell and M.~Schvellinger,
  ``On the Delta(I) = 1/2 rule in holographic QCD,''
  Phys.\ Rev.\  D {\bf 74}, 026003 (2006)
  [arXiv:hep-ph/0512089].
  %%CITATION = PHRVA,D74,026003;%%
  
 \bibitem{Schafer:2007qy}
  T.~Schafer,
  ``Euclidean Correlation Functions in a Holographic Model of QCD,''
  arXiv:0711.0236 [hep-ph].
  %%CITATION = ARXIV:0711.0236;%%

\bibitem{Csaki:2006ji}
  C.~Csaki and M.~Reece,
  ``Toward a systematic holographic QCD: A braneless approach,''
  JHEP {\bf 0705}, 062 (2007)
  [arXiv:hep-ph/0608266].
  %%CITATION = JHEPA,0705,062;%%
  
  \bibitem{Batell:2008zm}
  B.~Batell and T.~Gherghetta,
  ``Dynamical Soft-Wall AdS/QCD,''
  arXiv:0801.4383 [hep-ph].
  %%CITATION = ARXIV:0801.4383;%%

\bibitem{Polchinski:2002jw}
  J.~Polchinski and M.~J.~Strassler,
  ``Deep inelastic scattering and gauge/string duality,''
  JHEP {\bf 0305}, 012 (2003)
  [arXiv:hep-th/0209211].
  %%CITATION = JHEPA,0305,012;%%
  
  \bibitem{Hatta:2007he}
  Y.~Hatta, E.~Iancu and A.~H.~Mueller,
  ``Deep inelastic scattering at strong coupling from gauge/string duality :
  the saturation line,''
  arXiv:0710.2148 [hep-th].
  %%CITATION = ARXIV:0710.2148;%%
  
  \bibitem{BallonBayona:2007rs}
  C.~A.~Ballon Bayona, H.~Boschi-Filho and N.~R.~F.~Braga,
  ``Deep inelastic structure functions from supergravity at small x,''
  arXiv:0712.3530 [hep-th].
  %%CITATION = ARXIV:0712.3530;%%

\bibitem{Brower:2006ea}
  R.~C.~Brower, J.~Polchinski, M.~J.~Strassler and C.~I.~Tan,
  ``The Pomeron and Gauge/String Duality,''
  JHEP {\bf 0712}, 005 (2007)
  [arXiv:hep-th/0603115].
  %%CITATION = JHEPA,0712,005;%%

\bibitem{Alday:2007hr}
  L.~F.~Alday and J.~M.~Maldacena,
  ``Gluon scattering amplitudes at strong coupling,''
  JHEP {\bf 0706}, 064 (2007)
  [arXiv:0705.0303 [hep-th]].
  %%CITATION = JHEPA,0706,064;%%
  
\bibitem{Karch:2002sh}
  A.~Karch and E.~Katz,
  ``Adding flavor to AdS/CFT,''
  JHEP {\bf 0206}, 043 (2002)
  [arXiv:hep-th/0205236].
  %%CITATION = JHEPA,0206,043;%%
  
  \bibitem{Kruczenski:2003be}
  M.~Kruczenski, D.~Mateos, R.~C.~Myers and D.~J.~Winters,
  ``Meson spectroscopy in AdS/CFT with flavour,''
  JHEP {\bf 0307} (2003) 049
  [arXiv:hep-th/0304032];
  %%CITATION = JHEPA,0307,049;%%
  %bibitem{Kruczenski:2003uq}
  %M.~Kruczenski, D.~Mateos, R.~C.~Myers and D.~J.~Winters,
  ``Towards a holographic dual of large-N(c) QCD,''
  JHEP {\bf 0405}, 041 (2004)
  [arXiv:hep-th/0311270].
  %%CITATION = JHEPA,0405,041;%%
    
  \bibitem{Babington:2003vm}
  J.~Babington, J.~Erdmenger, N.~J.~Evans, Z.~Guralnik and I.~Kirsch,
  ``Chiral symmetry breaking and pions in non-supersymmetric gauge /  gravity
  duals,''
  Phys.\ Rev.\  D {\bf 69}, 066007 (2004)
  [arXiv:hep-th/0306018].
  %%CITATION = PHRVA,D69,066007;%%
  
   \bibitem{Sakai:2004cn}
  T.~Sakai and S.~Sugimoto,
  ``Low energy hadron physics in holographic QCD,''
  Prog.\ Theor.\ Phys.\  {\bf 113}, 843 (2005)
  [arXiv:hep-th/0412141];
  %%CITATION = PTPKA,113,843;%%
  % \bibitem{Sakai:2005yt}
  T.~Sakai and S.~Sugimoto,
  ``More on a holographic dual of QCD,''
  Prog.\ Theor.\ Phys.\  {\bf 114}, 1083 (2006)
  [arXiv:hep-th/0507073].
  %%CITATION = PTPKA,114,1083;%%
  
  \bibitem{Gursoy:2007er}
  U.~Gursoy, E.~Kiritsis and F.~Nitti,
  ``Exploring improved holographic theories for QCD: Part II,''
  arXiv:0707.1349 [hep-th].
  %%CITATION = ARXIV:0707.1349;%%

  \bibitem{Erdmenger:2007cm}
  J.~Erdmenger, N.~Evans, I.~Kirsch and E.~Threlfall,
  ``Mesons in Gauge/Gravity Duals - A Review,''
  arXiv:0711.4467 [hep-th].
  %%CITATION = ARXIV:0711.4467;%%
 
\bibitem{Dirac:1949cp}
  P.~A.~M.~Dirac,
  ``Forms of Relativistic Dynamics,''
  Rev.\ Mod.\ Phys.\  {\bf 21}, 392 (1949).
  %%CITATION = RMPHA,21,392;%%
  
  \bibitem{Balasubramanian:1998sn}
  V.~Balasubramanian, P.~Kraus and A.~E.~Lawrence,
  ``Bulk vs. boundary dynamics in anti-de Sitter spacetime,''
  Phys.\ Rev.\  D {\bf 59}, 046003 (1999)
  [arXiv:hep-th/9805171].
  %%CITATION = PHRVA,D59,046003;%%
  
  \bibitem{Klebanov:1999tb}
  I.~R.~Klebanov and E.~Witten,
  ``AdS/CFT correspondence and symmetry breaking,''
  Nucl.\ Phys.\  B {\bf 556}, 89 (1999)
  [arXiv:hep-th/9905104].
  %%CITATION = NUPHA,B556,89;%%
  
    \bibitem{Breitenlohner:1982jf}
  P.~Breitenlohner and D.~Z.~Freedman,
  ``Stability In Gauged Extended Supergravity,''
  Annals Phys.\  {\bf 144}, 249 (1982).
  %%CITATION = APNYA,144,249;%%
  
\bibitem{l'Yi:1998eu}
  W.~S.~l'Yi,
   ``Correlators of currents corresponding to the massive p-form fields in
  AdS/CFT correspondence,''
  Phys.\ Lett.\ B {\bf 448}, 218 (1999)
  [arXiv:hep-th/9811097].
  %%CITATION = HEP-TH 9811097;%%
  
  \bibitem{Brodsky:1997de}
  S.~J.~Brodsky, H.~C.~Pauli and S.~S.~Pinsky,
  ``Quantum chromodynamics and other field theories on the light cone,''
  Phys.\ Rept.\  {\bf 301}, 299 (1998)
  [arXiv:hep-ph/9705477].
  %%CITATION = PRPLC,301,299;%%
  
  \bibitem{Yao:2006px}
  W.~M.~Yao {\it et al.}  [Particle Data Group],
  ``Review of particle physics,''
  J.\ Phys.\ G {\bf 33}, 1 (2006).
  %%CITATION = JPHGB,G33,1;%%
  
  \bibitem{Infeld:1941}
  L. Infeld,
  ``On a New Treatment of Some Eigenvalue Problems'',
  Phys.\ Rev. {\bf 59}, 737 (1941).

 \bibitem{Metsaev:1999kb}
  R.~R.~Metsaev,
  ``IIB supergravity and various aspects of light-cone formalism in AdS
  space-time,''
  arXiv:hep-th/0002008.
  %%CITATION = HEP-TH/0002008;%%
  
  \bibitem{Andreev:2006vy}
  O.~Andreev,
  ``1/q**2 corrections and gauge / string duality,''
  Phys.\ Rev.\  D {\bf 73}, 107901 (2006)
  [arXiv:hep-th/0603170].
  %%CITATION = PHRVA,D73,107901;%%
  
    \bibitem{deTeramond:2006xb}
  G.~F.~de Teramond,
  ``Mapping string states into partons: Form factors and the hadron spectrum in
  AdS/QCD,''
  arXiv:hep-ph/0606143.
  %%CITATION = HEP-PH/0606143;%%

\bibitem{Radyushkin:2006iz}
  A.~V.~Radyushkin,
  ``Holographic wave functions, meromorphization and counting rules,''
  Phys.\ Lett.\  B {\bf 642}, 459 (2006)
  [arXiv:hep-ph/0605116].
  %%CITATION = PHLTA,B642,459;%%

\bibitem{Grigoryan:2007vg}
  H.~R.~Grigoryan and A.~V.~Radyushkin,
  ``Form Factors and Wave Functions of Vector Mesons in Holographic QCD,''
  Phys.\ Lett.\  B {\bf 650}, 421 (2007)
  [arXiv:hep-ph/0703069];
  %%CITATION = PHLTA,B650,421;%%

\bibitem{Grigoryan:2007my}
  H.~R.~Grigoryan and A.~V.~Radyushkin,
  ``Structure of Vector Mesons in Holographic Model with Linear Confinement,''
  Phys.\ Rev.\  D {\bf 76}, 095007 (2007)
  [arXiv:0706.1543 [hep-ph]].
  %%CITATION = PHRVA,D76,095007;%%
  
  \bibitem{Brodsky:2007hb}
  S.~J.~Brodsky and G.~F.~de Teramond,
  ``Light-Front Dynamics and AdS/QCD: The Pion Form Factor in the Space- and
  Time-Like Regions,''
  arXiv:0707.3859 [hep-ph].
  %%CITATION = ARXIV:0707.3859;%%
  
  \bibitem{Brodsky:2007vk}
  S.~J.~Brodsky and G.~F.~de Teramond,
  ``AdS/CFT and Exclusive Processes in QCD,''
  arXiv:0709.2072 [hep-ph].
  %%CITATION = ARXIV:0709.2072;%%
  
  \bibitem{Huang:2007uu}
  T.~Huang and F.~Zuo,
  ``Couplings of the Rho Meson in a Holographic dual of QCD with Regge
  Trajectories,''
  arXiv:0708.0936 [hep-ph].
  %%CITATION = ARXIV:0708.0936;%%

\bibitem{Kwee:2007dd}
  H.~J.~Kwee and R.~F.~Lebed,
  ``Pion Form Factors in Holographic QCD,''
  arXiv:0708.4054 [hep-ph].
  %%CITATION = ARXIV:0708.4054;%%

 \bibitem{Abidin:2008ku}
  Z.~Abidin and C.~E.~Carlson,
  ``Gravitational Form Factors of Vector Mesons in an AdS/QCD Model,''
  arXiv:0801.3839 [hep-ph].
  %%CITATION = ARXIV:0801.3839;%%

\bibitem{Hong:2004sa}
  S.~Hong, S.~Yoon and M.~J.~Strassler,
  ``On the couplings of vector mesons in AdS/QCD,''
  JHEP {\bf 0604}, 003 (2006)
  [arXiv:hep-th/0409118].
  %%CITATION = JHEPA,0604,003;%%
  
  \bibitem{Polchinski:2001ju}
  J.~Polchinski and L.~Susskind,
  ``String theory and the size of hadrons,''
  arXiv:hep-th/0112204.
  %%CITATION = HEP-TH/0112204;%%
  
  \bibitem{Baldini:1998qn}
  R.~Baldini, S.~Dubnicka, P.~Gauzzi, S.~Pacetti, E.~Pasqualucci and Y.~Srivastava,
  ``Nucleon time-like form factors below the N anti-N threshold,''
  Eur.\ Phys.\ J.\  C {\bf 11}, 709 (1999).
  %%CITATION = EPHJA,C11,709;%%

\bibitem{Tadevosyan:2007yd}
  V.~Tadevosyan {\it et al.}  [Jefferson Lab F(pi) Collaboration],
  ``Determination of the pion charge form factor for $Q^2=0.60-1.60 {\rm GeV}^2$,''
  Phys.\ Rev.\  C {\bf 75}, 055205 (2007)
  [arXiv:nucl-ex/0607007];
  %%CITATION = PHRVA,C75,055205;%%
%\bibitem{Horn:2006tm}
  T.~Horn {\it et al.}  [Jefferson Lab F(pi)-2 Collaboration],
  ``Determination of the Charged Pion Form Factor at $Q^2=1.60$  and $2.45
  ({\rm GeV/c})^2$,''
  Phys.\ Rev.\ Lett.\  {\bf 97}, 192001 (2006)
  [arXiv:nucl-ex/0607005].
  %%CITATION = PRLTA,97,192001;%%

 \bibitem{Diehl:2005wq}
  M.~Diehl,
  ``Generalized parton distributions from form factors,''
  Nucl.\ Phys.\ Proc.\ Suppl.\  {\bf 161}, 49 (2006)
  [arXiv:hep-ph/0510221].
  %%CITATION = NUPHZ,161,49;%%

  \bibitem{Pauli:1985ps}
  H.~C.~Pauli and S.~J.~Brodsky,
  ``Discretized Light Cone Quantization: Solution To A Field Theory In One
  Space One Time Dimensions,''
  Phys.\ Rev.\  D {\bf 32}, 2001 (1985).
  %%CITATION = PHRVA,D32,2001;%%

\bibitem{Brodsky:2000ii}
  S.~J.~Brodsky, D.~S.~Hwang, B.~Q.~Ma and I.~Schmidt,
  ``Light-cone representation of the spin and orbital angular momentum of
  relativistic composite systems,''
  Nucl.\ Phys.\  B {\bf 593}, 311 (2001)
  [arXiv:hep-th/0003082].
  %%CITATION = NUPHA,B593,311;%%

\bibitem{Brodsky:1998hn}
  S.~J.~Brodsky and D.~S.~Hwang,
  ``Exact light-cone wavefunction representation of matrix elements of
  electroweak currents,''
  Nucl.\ Phys.\  B {\bf 543}, 239 (1999)
  [arXiv:hep-ph/9806358].
  %%CITATION = NUPHA,B543,239;%%
  
  \bibitem{Brodsky:2000xy}
  S.~J.~Brodsky, M.~Diehl and D.~S.~Hwang,
  ``Light-cone wavefunction representation of deeply virtual Compton
  scattering,''
  Nucl.\ Phys.\  B {\bf 596}, 99 (2001)
  [arXiv:hep-ph/0009254].
  %%CITATION = NUPHA,B596,99;%%
  

\bibitem{Berger:1979du}
  E.~L.~Berger and S.~J.~Brodsky,
  %``Quark Structure Functions Of Mesons And The Drell-Yan Process,''
  Phys.\ Rev.\ Lett.\  {\bf 42}, 940 (1979).
  %%CITATION = PRLTA,42,940;%%
  
  \bibitem{Brodsky:2008}
S.~J.~Brodsky, `` Evidence for Color Transparency and Direct Hadron Production at RHIC"
RHIC News, January 15, 2008.

  \bibitem{Lepage:1979zb}
  G.~P.~Lepage and S.~J.~Brodsky,
  ``Exclusive Processes In Quantum Chromodynamics: Evolution Equations For
  Hadronic Wave Functions And The Form-Factors Of Mesons,''
  Phys.\ Lett.\  B {\bf 87}, 359 (1979).
  %%CITATION = PHLTA,B87,359;%%
 

\bibitem{Efremov:1979qk}
  A.~V.~Efremov and A.~V.~Radyushkin,
  ``Factorization And Asymptotical Behavior Of Pion Form-Factor In QCD,''
  Phys.\ Lett.\  B {\bf 94}, 245 (1980).
  %%CITATION = PHLTA,B94,245;%%
  


\bibitem{Brodsky:1979qm}
  S.~J.~Brodsky and G.~P.~Lepage,
  ``Exclusive Processes And The Exclusive Inclusive Connection In Quantum
  Chromodynamics,''
  %%CITATION = SLAC-PUB-2294;%%
  SLAC-PUB-2294 (1979).
  Presented at  the Workshop on Current Topics in High Energy Physics, Cal Tech. (1979).

 \bibitem{Drell:1969km}
  S.~D.~Drell and T.~M.~Yan,
  ``Connection Of Elastic Electromagnetic Nucleon Form-Factors At Large $Q^2$
  And Deep Inelastic Structure Functions Near Threshold,''
  Phys.\ Rev.\ Lett.\  {\bf 24}, 181 (1970).
  %%CITATION = PRLTA,24,181;%%

\bibitem{West:1970av}
  G.~B.~West,
  ``Phenomenological model for the electromagnetic structure of the proton,''
  Phys.\ Rev.\ Lett.\  {\bf 24}, 1206 (1970).
  %%CITATION = PRLTA,24,1206;%%

\bibitem{Brodsky:1980zm}
  S.~J.~Brodsky and S.~D.~Drell,
  ``The Anomalous Magnetic Moment And Limits On Fermion Substructure,''
  Phys.\ Rev.\  D {\bf 22}, 2236 (1980).
  %%CITATION = PHRVA,D22,2236;%%

\bibitem{Donnellan:2007xr}
  M.~A.~Donnellan {\it et al.},
  ``Lattice Results for Vector Meson Couplings and Parton Distribution
  Amplitudes,''
  arXiv:0710.0869 [hep-lat].
  %%CITATION = NONE,,;%%

\bibitem{Braun:2006dg}
  V.~M.~Braun {\it et al.},
  ``Moments of pseudoscalar meson distribution amplitudes from the lattice,''
  Phys.\ Rev.\  D {\bf 74}, 074501 (2006)
  [arXiv:hep-lat/0606012].
  %%CITATION = PHRVA,D74,074501;%%

\bibitem{Aitala:2000hb}
  E.~M.~Aitala {\it et al.}  [E791 Collaboration],
  ``Direct measurement of the pion valence quark momentum distribution, the
  pion light-cone wave function squared,''
  Phys.\ Rev.\ Lett.\  {\bf 86}, 4768 (2001)
  [arXiv:hep-ex/0010043].
  %%CITATION = PRLTA,86,4768;%%

\bibitem{Choi:2006ha}
  H.~M.~Choi and C.~R.~Ji,
  ``Conformal Symmetry and Pion Form Factor: Soft and Hard Contributions,''
  Phys.\ Rev.\  D {\bf 74}, 093010 (2006)
  [arXiv:hep-ph/0608148].
  %%CITATION = PHRVA,D74,093010;%%
  
   \bibitem{Brodsky:2006uq}
  S.~J.~Brodsky,
  ``Hadron Spectroscopy and Structure from AdS/CFT,''
  Eur.\ Phys.\ J.\  A {\bf 31}, 638 (2007)
  [arXiv:hep-ph/0610115].
  %%CITATION = EPHJA,A31,638;%%

 \bibitem{Brodsky:2005kc}
  S.~J.~Brodsky and G.~F.~de Teramond,
  ``Hadron spectroscopy and wavefunctions in QCD and the AdS/CFT
  correspondence,''
  AIP Conf.\ Proc.\  {\bf 814}, 108 (2006)
  [arXiv:hep-ph/0510240].
  %%CITATION = APCPC,814,108;%%

\bibitem{Burkardt:2005td}
  M.~Burkardt,
  ``Hadron tomography,''
  Int.\ J.\ Mod.\ Phys.\  A {\bf 21}, 926 (2006)
  [arXiv:hep-ph/0509316].
  %%CITATION = IMPAE,A21,926;%%

\bibitem{Ji:2003ak}
  X.~d.~Ji,
  ``Viewing the proton through 'color'-filters,''
  Phys.\ Rev.\ Lett.\  {\bf 91}, 062001 (2003)
  [arXiv:hep-ph/0304037].
  %%CITATION = PRLTA,91,062001;%%
  
  \bibitem{Ralston:2001xs}
  J.~P.~Ralston and B.~Pire,
  ``Femto-photography of protons to nuclei with deeply virtual Compton
  scattering,''
  Phys.\ Rev.\  D {\bf 66}, 111501 (2002)
  [arXiv:hep-ph/0110075].
  %%CITATION = PHRVA,D66,111501;%%

\bibitem{Brodsky:2006in}
  S.~J.~Brodsky, D.~Chakrabarti, A.~Harindranath, A.~Mukherjee and J.~P.~Vary,
  ``Hadron optics: Diffraction patterns in deeply virtual Compton scattering,''
  Phys.\ Lett.\  B {\bf 641}, 440 (2006)
  [arXiv:hep-ph/0604262].
  %%CITATION = PHLTA,B641,440;%%

\bibitem{Hoyer:2006xg}
  P.~Hoyer,
  ``Comments on the relativity of shape,''
  AIP Conf.\ Proc.\  {\bf 904}, 65 (2007)
  [arXiv:hep-ph/0608295].
  %%CITATION = APCPC,904,65;%%

\bibitem{Brodsky:2006ku}
  S.~J.~Brodsky, D.~Chakrabarti, A.~Harindranath, A.~Mukherjee and J.~P.~Vary,
  ``Hadron optics in three-dimensional invariant coordinate space from deeply
  virtual Compton scattering,''
  Phys.\ Rev.\  D {\bf 75}, 014003 (2007)
  [arXiv:hep-ph/0611159].
  %%CITATION = PHRVA,D75,014003;%%


 \end{thebibliography}
\end{document}